\documentclass[useAMS,usenatbib]{mn2e/mn2e}
\usepackage{graphicx}
\usepackage{txfonts}
\usepackage{url}
\usepackage{color}
\usepackage{enumitem}
\usepackage{verbatim}
\usepackage{footnote}
\usepackage[export]{adjustbox}
\makesavenoteenv{tabular}

\newcommand{\jks}{\mbox{$J\!-\!K_{\rm s}$}}
\newcommand{\ks}{\mbox{$K_{\rm s}$}}

\title[Dust production rate of carbon stars in the SMC]{Estimating the dust production rate of carbon stars in the Small Magellanic Cloud}
\author[Nanni et al.]{Ambra Nanni$^1$,
Paola Marigo$^1$, L\'eo Girardi$^2$, Stefano Rubele$^1$, 
Alessandro Bressan$^3$, \newauthor
Martin A.T. Groenewegen$^4$,
Giada Pastorelli$^1$,
Bernhard Aringer$^1$
  \\
  $^1$ Dipartimento di Fisica e Astronomia Galileo Galilei,
  Universit\`a di Padova, Vicolo dell'Osservatorio 3, I-35122 Padova, Italy\\
  $^2$ Osservatorio Astronomico di Padova, Vicolo dell'Osservatorio 5,
  I-35122 Padova, Italy \\
  $^3$ SISSA, via Bonomea 265, I-34136 Trieste, Italy\\
  $^4$ Koninklijke Sterrenwacht van Belgi\"e, Ringlaan 3, B-1180 Brussel, Belgium  \\
}

\begin{document}

\date{Accepted .  Received ; in original form }

\pagerange{\pageref{firstpage}--\pageref{lastpage}} \pubyear{2017}

\maketitle

\begin{abstract}\label{firstpage}
We employ newly computed grids of spectra reprocessed by dust for estimating the total dust production rate (DPR) of carbon stars in the Small Magellanic Cloud (SMC).
For the first time, the grids of spectra are computed as a function of the main stellar parameters, i.e. mass-loss rate, luminosity, effective temperature, current stellar mass and element abundances at the photosphere, following a consistent, physically grounded scheme of dust growth coupled with stationary wind outflow. The model accounts for the dust growth of various dust species formed in the circumstellar envelopes of carbon stars, such as carbon dust, silicon carbide and metallic iron. In particular, we employ some selected combinations of optical constants and grain sizes for carbon dust which have been shown to reproduce simultaneously the most relevant color-color diagrams in the SMC. By employing our grids of models, we fit the spectral energy distributions of $\approx$3100 carbon stars in the SMC, consistently deriving some important dust and stellar properties, i.e. luminosities, mass-loss rates, gas-to-dust ratios, expansion velocities and dust chemistry. We discuss these properties and we compare some of them with observations in the Galaxy and LMC. 
We compute the DPR of carbon stars in the SMC, finding that the estimates provided by our method can be significantly different, between a factor $\approx2-5$, than the  ones available in the literature. 

Our grids of models, including the spectra and other relevant dust and stellar quantities, are publicly available at \url{http://starkey.astro.unipd.it/web/guest/dustymodels}.

\end{abstract}


\section{Introduction}
\label{introduction}
Carbon-rich (C) stars on the thermally pulsing asymptotic giant branch (TP-AGB) are characterized by luminous and cool atmospheres with plenty of spectral features from C-bearing molecules, and by extended dust-rich circumstellar envelopes (CSE). Dust grains can deeply modify the emerging spectral energy distributions (SED) of such stars, because they absorb the photospheric stellar radiation and re-emit it at longer wavelength. In addition to that, photons are also scattered by dust grains. As a consequence, C-rich stars are extremely relevant for the interpretation of the near- and mid-infrared colors (NIR and MIR) of both resolved and unresolved stellar populations. This is especially true at the lower-than-solar metallicities that characterize most dwarf galaxies, for which a large fraction of the TP-AGB stars evolves through a C-rich phase.

Indeed, galaxies such as the Magellanic Clouds contain rich population of C stars which can be classified roughly in three groups: (1) those belonging to the ``red tail'' of C stars, with \jks\ colors between $1.2$ and $\approx2$~mag \citep{Cioni06}; (2) the dust-obscured sources classified as ``extreme AGB-stars'' (X-stars), extending to much redder \jks\ colors, which are most likely to be C-rich \citep{vanLoon97, vanLoon06, vanLoon08, Matsuura09}, and finally (3) C-rich members among the class dubbed ``anomalous O-rich AGB stars'' (hereafter aAGB) by \citet{Boyer11}, which have \jks\ similar to those of O-rich stars but redder $J-[8]$ colors. Their infrared SEDs can be derived, for a wide range of wavelengths, from high-quality photometric surveys such as the Two Micron All Sky Survey \citep[2MASS;][]{skrutskie06}, and the Spitzer surveys of the Large Magellanic Cloud (LMC) \citep[SAGE;][]{blum06} and Small Magellanic Cloud (SMC) \citep[S$^{3}$MC;][]{bolatto07}. 
As far as the SMC is concerned, the most comprehensive set of infrared data in the range between 3.6-160~$\mu$m is presented in the Spitzer Space Telescope Legacy Program ``Surveying the Agents of Galaxy Evolution in the tidally stripped, low metallicity SMC'' \citep[SAGE-SMC,][]{gordon11}, from which about 5800 TP-AGB stars have been classified \citep{Boyer11,Boyer15}. A revised version of the original catalog by \citet{Boyer11} is presented in the TP-AGB candidate list by \citet{Srinivasan16}.

In the latest years, different groups aimed at estimating the total dust production rate (DPR) and the individual mass-loss rates of AGB stars in the MCs employing the SED fitting technique \citep{vanLoon06_2,Groenewegen07, Groenewegen09, Srinivasan11, Gullieuszik12, Boyer12, Riebel12, Matsuura13, Srinivasan16,Goldman17}.
Pre-computed grids of spectra for dusty AGB stars, employed for the SED fitting procedure, are available in the literature \citep{Groenewegen06,Srinivasan11}.
In the standard approach adopted in the literature, the SEDs of AGB stars are fitted by choosing a priori a given optical data set and a certain grain size or grain size distribution. 
However, the typical grain size of carbon dust is uncertain and several optical data sets for carbon dust, very different from each other, are available in the literature \citep{Hanner88,Rouleau91,Zubko96,Jaeger98}. 
As discussed in \citet{Nanni16} both the optical data set and grain size heavily affect the emerging spectra of dust-enshrouded C-rich stars and the carbon dust optical constants need to be constrained by reproducing most of the NIR and MIR color-color diagrams (CCDs) simultaneously. In the context of hydrodynamical models, optical constants and grain sizes of carbon dust have been also discussed by \citet{Andersen99}.
In addition to that, the SED fitting technique relies on assumptions related to dust chemistry, gas-to-dust ratio (usually assumed to be fixed), outflow expansion velocity, dust condensation temperature, and shell geometry. The evaluation of the dust mass-loss rate is also sensitive to the dust temperature at the boundary of the inner shell, which is usually fixed, and to the typical size of dust grains, which also needs to be assumed. 

In this work, we present a new grid of dusty models based on a physically grounded scheme for dust growth, coupled with a stationary wind \citep{Nanni13, Nanni14}. 
By using such a grid of models, we fit the SEDs for all the carbon stars in the catalog of the SMC by \citet{Srinivasan16}, and we compute their total DPR.
In contrast to the standard approach used so far in the literature, our dust model consistently computes the gas-to-dust ratio, dust chemistry and outflow expansion velocity, without the need of relying on assumptions or scaling relations for these quantities. 
We also investigate the variations on the final dust budget produced by employing different optical data sets and grain sizes for the SED fitting. Such combinations are the ones which reproduce the most importan NIR and MIR colors simultaneously in the SMC and have been selected in \citet{Nanni16}.

\section{Model and SED fitting method}\label{model}
In this Section we recall the basic equations of our dust growth model and discuss the sample of carbon-rich stars fitted as well as the SED fitting method.

\subsection{Dust growth scheme}
We adopt the dust growth scheme described in \citet{Nanni13, Nanni14} which is an improved version of the description by \citet{FG06} also employed, in its original formulation, by other groups \citep{Ventura12,Ventura14, Ventura16,Dellagli15b, Dellagli15a}.
Our code requires as input quantities the stellar parameters (luminosity, $L$, effective temperature, $T_{\rm eff}$, photospheric spectrum, actual stellar mass, $M$, element abundances in the atmosphere and mass-loss rate $\dot{M}$), plus the seed particle abundance in the CSEs, the optical data set of the dust species, and the initial conditions, which are the initial grain size $a_0=10^{-3}$ $\mu$m, which is assumed to be the same for all the dust species and the initial outflow velocity $v_{\rm i}= 4$ km s$^{-1}$. In case the outflow is not accelerated, the dust production is computed assuming a constant value of the velocity $v_{\rm exp}=v_{\rm i}$. The value of the initial velocity is selected in order to reproduced the observed C- and X-stars in stellar population synthesis models (Pastorelli et al., in preparation).

We include as dust species amorphous carbon (amC), silicon carbide (SiC) and metallic iron. 
For amorphous carbon we select the combinations of optical data sets and grain sizes which well reproduce several of the observed color-color diagrams in the NIR and MIR bands, on the base of \citet{Nanni16}. Such combinations are listed in Table~\ref{opt}.
The optical data sets for SiC and metallic iron are taken from \citet{Pegourie88} and \citet{Leksina67}, respectively.

The seed abundance is assumed to be proportional to the carbon-excess $\epsilon_{\rm C}-\epsilon_{\rm O}$:
\begin{equation}\label{n_seeds}
\epsilon_{\rm s, C}\propto\epsilon_{\rm s} (\epsilon_{\rm C}-\epsilon_{\rm O}),
\end{equation}
where $\epsilon_{\rm s}$ is a free model parameter \citep{Nanni16}. 
The quantities $\epsilon_{\rm C}$ and $\epsilon_{\rm O}$ are the number densities of carbon and oxygen atoms in the stellar atmosphere normalized by the number of hydrogen nuclei.
The quantity $\epsilon_{\rm s, C}$
is assumed to be the same for all the dust species formed.

Given the input quantities, the code integrates a set of differential equations describing the dust growth of various dust species, the stationary, spherically symmetric, outflow and the envelope structure.

\begin{table*}
\begin{center}
\caption{Combination of optical data sets and seed particle abundances selected for the SED fitting.}
\label{opt}
\begin{tabular}{l c c c}
\hline
Optical data set  & $\rho_{\rm d, amC}$ [g cm$^{-3}$]& $\log(\epsilon_{\rm s})$ &  Denomination \\
\hline
\citet{Rouleau91} & 1.85 & $-12$ & R12 \\
\citet{Rouleau91}  & 1.85 & $-13$ & R13 \\
\citet{Jaeger98} (T=400 $^\circ$C)  & 1.435 &$-12$ &  J400 \\
\citet{Jaeger98} (T=1000 $^\circ$C) & 1.988 &$-12$   &  J1000 \\
\citet{Hanner88} &  1.85 & $-11$ &   H11 \\
\citet{Zubko96} (ACAR sample) & 1.87 &$-12$ &  Z12\\
\hline
\end{tabular}
\end{center}
\end{table*}

For any combination of the stellar quantities, our dust scheme provides as output the outflow structure in terms of density profile, outflow velocity, dust condensation radius and dust properties, such as the chemical composition, the gas-to-dust ratio and the dust temperature at the boundary of the inner shell. 

The photospheric spectra, taken from \citet{Aringer16}, are reprocessed by dust. The radiative transfer calculation is performed by means of the code More of \textsc{dusty} \citep[MoD;][]{Groenewegen12}, based on \textsc{dusty} \citep{Ivezic97}. Some of the quantities computed by our dust formation code, as the optical depth at a given fiducial wavelength, $\tau_\lambda$, the average optical properties consistently computed for the chemistry and grain size of the different dust species and the dust temperature at the inner boundary of the shell, are taken as input in MoD.

We briefly recall in the following the most useful equations describing the dust growth process in CSEs \citep{Nanni13, Nanni14}.

For the dust growth calculation along the CSE, the basic equations are the following.
\begin{itemize}
\item Gas temperature profile:
\begin{equation}\label{gas_temp}
T_{\rm gas}(r)^4=T_{\rm eff}^4\left[W(r)+\frac{3}{4}\tau_{\rm L}\right],
\end{equation}
where $W(r)$ is the dilution term:
\begin{equation}
W(r)=\frac{1}{2}\left[1-\sqrt{1-\left(\frac{R_*}{r}\right)^2}\right],
\end{equation}
and $\tau_{\rm L}$ is defined by the differential equation:
\begin{equation}\label{tauL}
\frac{d\tau_{\rm L}}{dr}=-\rho\kappa \left(\frac{R_*}{r}\right)^2,
\end{equation}
where $\kappa$ is the average opacity of the medium computed as fully described in \citet{Nanni13, Nanni14}, $R_*$ is the stellar radius, $r$ is the distance from the center of the star and the quantity $\tau_{\rm L}$ has to be zero at infinity. 
Note that the temperature structure determines by Eq.~\ref{gas_temp} is dependent on the amount of dust produced through the term $\tau_{\rm L}$, which contains the quantity $\kappa$.

\item Growth rate of the dust grain radius.\\
Once the seed particle abundance is assumed, the grain growth proceeds through the addition of molecules on the grain surface. The differential equation which describes such a process is:
\begin{equation}\label{dadt}
 \frac{da_{\rm i}}{dt}=V_{\rm 0,i} (J^{\rm gr}_{\rm i}-J^{\rm dec}_{\rm i}),
\end{equation}
where $J^{\rm gr}_{\rm i}$ and $J^{\rm dec}_{\rm i}$ are the growth and decomposition rates, respectively and $V_{\rm 0, i}$ is the volume of one monomer of dust. The term $J^{\rm gr}_{\rm i}$ is provided by the rate of effective collisions of the molecules impinging on the grain surface, while $J^{\rm dec}_{\rm i}$ can be provided by pure sublimation of the dust grains due to heating from the stellar radiation and/or by the inverse reaction between H$_2$ molecules at the grain surface (chemisputtering).

For carbon dust the chemisputtering term is assumed to be negligible following the scheme by \citet{Cherchneff92} in which carbon dust accretes below a certain threshold gas temperature, $T_{\rm gas}=1100$~K.  For such a gas temperature the sublimation process of carbon grains is usually not at work.  Iron dust does not react with H$_2$ molecules and only sublimation is included in the decomposition term for this dust species. For SiC chemisputtering is more efficient than pure sublimation and is included in the decomposition term.

We define the condensation radius of a certain dust species, $i$, $R_{\rm c, i}$, as the distance at which  $J^{\rm gr}_{\rm i} \ge J^{\rm dec}_{\rm i}$.
In the case of carbon dust, the condensation radius is depending on the gas temperature profile which is also dependent on the amount of dust produced through Eq.~\ref{tauL}.

\item Expansion velocity profile for a spherically symmetric, stationary outflow
\begin{equation}\label{velocity}
v \frac{dv}{dt}=-\frac{G M}{r^2}(1-\Gamma),
\end{equation}
where $G$ is the gravitational constant and the quantity
\begin{equation}\label{gamma}
\Gamma=\frac{L}{4\pi c G M} \kappa,
\end{equation}
is the ratio between the radiation pressure and the gravitational pull of the star. The constant $c$ is the speed of light.

Eqs.~\ref{tauL}, \ref{dadt} (one equation for each dust species) and \ref{velocity} provide the complete set of differential equations to integrate.

\item Gas profile, described by the mass conservation equation:
\begin{equation}\label{dens}
\rho=\frac{\dot{M}}{4\pi r^2 v},
\end{equation}
where the velocity, $v$, changes along the CSE according to Eq.~\ref{velocity}. 

\item The dust-to-gas ratio, $\delta_{\rm i}$, for any of the dust species, $i$:
\begin{equation}\label{delta_i}
\delta_i=\frac{X_{\rm k, i}}{m_{\rm k,i}}f_{\rm i}\frac{m_{\rm i}}{n_{\rm k,i}}
\end{equation}
where $X_{\rm k, i}$, is the mass fraction of the key-element\footnote{The least abundant of the elements in the stellar atmosphere forming a given type of dust.}, $m_{\rm k,i}$ is its atomic mass, $f_{\rm i}$ is the number fraction of condensed key-element particles over the total, $n_{\rm k,i}$ is the number of atoms of the key-element in one monomer of dust and $m_{\rm i}$ is the mass of the monomer.
The quantity $f_{\rm i}$ for each dust species is computed following the grain growth in Eq.~\ref{dadt} and it changes along the CSEs.

\item Dust density profile, derived from Eq.~\ref{dens}:
\begin{equation}\label{d_dens}
\rho_{\rm dust}=\rho\sum_i\delta_{\rm i},
\end{equation}

\item Dust mass-loss rate, given by the contribution of all the dust species, $i$,
\begin{equation}\label{d_mloss}
\dot{M}_{\rm dust}=\sum_i \dot{M}_i=\dot{M}\times\delta_i.
\end{equation}
\item Total gas-to-dust ratio $\Psi_{\rm dust}$, given by the ratio between the gas and dust mass-loss rates:
\begin{equation}\label{psi_dust}
\Psi_{\rm dust}=\frac{\dot{M}}{\dot{M}_{\rm dust}},
\end{equation}
\end{itemize}
\vskip 3mm
For the a posteriori radiative transfer calculation, the input quantities are the following.
\begin{itemize}
\item The optical depth at a given wavelength is computed as
\begin{equation}\label{tau_lambda}
\tau_\lambda=\frac{3\dot{M}}{4}\int_{R_{\rm c}}^{\infty} \sum_i \frac{Q_{\rm ext, i}(\lambda,a_i)}{a_i \rho_{\rm i}}\frac{\delta_i(r)}{r^2 v(r)}dr.
\end{equation}
where $R_{\rm c}$ is the condensation radius of the first dust species condensed, expressed in units of stellar radii R$_*$, and $\dot{M}$ is assumed to be constant.
The quantity $Q_{\rm ext}$ is the dust extinction coefficient, defined as:
\begin{equation}
Q_{\rm ext}(a_{\rm i}, \lambda)= Q_{\rm abs, i}(a_{\rm i},\lambda)+ Q_{\rm sca, i}(a_{\rm i}, \lambda) -g \times Q_{\rm sca, i}(a_{\rm i}, \lambda),
\end{equation}
where $g$ is defined as $g=$\,$<\cos\theta>$ and $\theta$ is the scattering angle. The quantity $g \times Q_{\rm sca, i}(a_{\rm i}, \lambda)$ provides the degree of forward scattering.

The quantities $Q_{\rm abs}(a_{\rm amC}, \lambda)$ and $Q_{\rm sca}(a_{\rm amC}, \lambda)$ are computed from the $n, k$ optical constants under the assumption of spherical dust grains by means of the code \textsc{bhmie} by \citet{Bohren83}, based on the Mie theory.

\item Dust temperature at the inner boundary of the shell, T$_{\rm inn}$.\\
The code MoD can only deal with a single temperature for all the dust species formed.
We take as representative temperature the carbon dust temperature at its condensation radius ($R_{\rm c, amC}$) plus one stellar radius, as explained in \citet{Nanni16}. Carbon dust in fact is usually the most abundant of the dust species formed in carbon-rich stars.
The carbon dust temperature is computed from the balance between the absorbed and emitted radiation.

\item Optical quantities representative for the dust mixture
\begin{equation}\label{av_Qext}
\bar{Q}_{\rm ext}=\frac{\sum_i \dot{M}_{\rm i} Q_{\rm ext, i}(a_{\rm i}, \lambda)}{\sum_i \dot{M}_{\rm i}},
\end{equation}
where $Q_{\rm ext, i}(a_{\rm i}, \lambda)$ is computed for the final grain size of each species, $i$, obtained by our dust formation model. 
\end{itemize}

\subsection{Grids of dusty models}
We select the six combinations of optical data sets and grain sizes for amorphous carbon dust listed in Table~\ref{opt} and discussed in \citet{Nanni16}. For SiC and metallic iron the optical data sets are fixed.
Different assumptions of optical data sets and grain sizes yield different inner temperatures for the dust zone and different values of optical depth at $\lambda=1$ $\mu$m, $\tau_1$, for the same input of the stellar quantities, because of different optical properties.

We then built six grids of models (one for each of the opacity set) by selecting a large range of stellar luminosity and mass-loss rates and some selected values of the actual stellar mass and carbon-excess, C$_{\rm ex}=\log(\epsilon_{\rm C}-\epsilon_{\rm O})+12$. The adopted metallicity is Z=0.004 with scaled solar abundances for the elements in the atmosphere (excluding carbon).

\begin{table}
\begin{center}
\caption{Input stellar parameters and spacing for the dust formation calculations.}
\label{Table:grid}
\begin{tabular}{l l l }
\hline
Parameter  &  Range/values  & spacing \\
\hline
$\log(L/L_\odot)$  & $[3.2,\,4]$  & 0.1   \\
                   & $[4.0,\, 4.4]$  & 0.05  \\
$\log(\dot{M}/M_{\odot} {\rm yr}^{-1})$    & $[-7,\, -5]$  & 0.1  \\
                   &  $[-5.0,\, -4.4]$ &  0.05  \\
$T_{\rm eff}$/K    &  $[2500,\,4000]$ & 100   \\
$M/M_{\odot}$     & 0.8, 1.5, 3  & \\
C$_{\rm ex}$ & 8.0, 8.2, 8.5  & \\
 C/O & 1.65, 2 , 3  & \\
Z &   0.004 &  \\
\hline
\end{tabular}
\end{center}
\end{table}

The range and spacing of the stellar parameters are summarized in Table~\ref{Table:grid}. They cover typical values for TP-AGB stars as predicted by stellar evolutionary models \citep{Marigo_etal17, marigoetal13}.
We compute the grids for all the effective temperatures available in the new grid of photospheric spectra presented in \citet{Aringer16}. The values of C/O have been interpolated in the grid of photospheric spectra for a metallicity value suitable for SMC stars.
We limit our calculation to spectra for which the combination of the stellar parameters produces models with $10^{-3}\le\tau_1\le 30 $.

The grids of dusty spectra and dust properties are publicly available at \url{http://starkey.astro.unipd.it/web/guest/dustymodels}. 

The data provided in the online grids of models consist on the input stellar quantities: 
\begin{itemize}
\item[-] Mass-loss rate, $\dot{M}$, in M$_\odot$yr$^{-1}$;
\item[-] Current stellar mass, $M$, in M$_\odot$;
\item[-] Effective temperature, T$_{\rm eff}$, in K;
\item[-] Stellar luminosity, $L$, in L$_\odot$;
\item[-] Carbon excess, C$_{\rm ex}$;
\item[-] C/O ratio;
\item[-] metallicity, $Z$, in mass fraction. 
\end{itemize}

The output quantities available for each combination of the input stellar parameters are:
\begin{itemize}
\item[-] Dust temperature at the inner boundary of the shell, $T_{\rm inn}$ in K;
\item[-] Mass-loss in dust for the different species, $\dot{M}_{\rm amC}$, $\dot{M}_{\rm SiC}$, $\dot{M}_{\rm iron}$ in  M$_\odot$yr$^{-1}$;
\item[-] outflow expansion velocity, $v_{\rm exp}$, in km~s$^{-1}$;
\item[-] optical depth at different wavelengths, $\tau_\lambda$;
\item[-] spectrum reprocessed by dust, normalized for the total luminosity.
\end{itemize}
As an example, we show in Table~\ref{Table:grid_example} the first lines of the online tables. The format of the files containing the spectra is the default one provided by the code \textsc{dusty} \citep{Ivezic97}.

\begin{table*}
\begin{center}
\caption{Example of input and output quantities contained in the publicly available online tables.}
\label{Table:grid_example}
\begin{tabular}{c c c c c c c c c c c c c c}
\hline
$\dot{M}$ & M & T$_{\rm eff}$ & L & C$_{\rm ex}$ & C/O & Z & T$_{\rm inn}$ & $\dot{M}_{\rm C}$ & $\dots$ & v$_{\rm exp}$ & $\tau_{\rm V}$ & $\cdots$   & spectrum \\
\hline
    3.16$\times10^{-7}$&      0.8&       2700&       15850&       8.5&       3.06&    0.004&       1209&   2.22$\times10^{-11}$ &   $\dots$  &       4.0 &     0.012& $\cdots$   &name1 \\
  3.98$\times10^{-7}$&      0.8&       2700&       15850&       8.5&       3.06&    0.004&       1211 &   5.30$\times10^{-11}$ &  $\dots$ &       4.2&     0.031&  $\cdots$    & name2 \\
   5.01$\times10^{-7}$ &      0.8&       2700&       15850&       8.5 &       3.06&    0.004&       1212&   1.02$\times10^{-10}$&  $\dots$ &       6.7&     0.063&  $\cdots$   &name3 \\
  6.31$\times10^{-7}$&      0.8&       2700&       15850&       8.5&       3.06&    0.004&       1212&   1.67$\times10^{-10}$&   $\dots$ &       9.5&      0.101& $\cdots$    &name4 \\
   7.943$\times10^{-7}$&      0.8 &       2700 &       15850 &       8.5 &      3.06 &    0.004 &       1212 &   2.53$\times10^{-10}$&  $\dots$ &       12.2 &      0.152 &  $\cdots$  & name5 \\
\hline
\end{tabular}
\end{center}
\end{table*}

\subsection{Comparison with previous approaches}
\label{comp_grams}

Without a description for dust formation, the optical depth is estimated from Eq.~\ref{tau_lambda} by assuming a) constant outflow velocity, $v_{\rm exp}$, b) fixed dust chemical composition for evaluating $\bar{Q}_{\rm ext}(\lambda, a)$ in Eq.~\ref{av_Qext}, c) same grain size (or grain size distribution) for all the dust species, d) a certain dust density, $\bar{\rho}_d$, averaged for all the dust species using $\dot{M}_{\rm i}$ as weight.
Under the above assumptions, Eq.~\ref{tau_lambda} reads as:
\begin{equation}\label{tau_approx}
\tau_\lambda{\rm (approx)}=\frac{3\dot{M}_{\rm dust}\bar{Q}_{\rm ext}(\lambda, a)}{4 a R_{\rm c} v_{\rm exp} \bar{\rho}_d}.
\end{equation}

The total DPR for each of the sources, $\dot{M}_{\rm dust}$, is then evaluated inverting Eq.~\ref{tau_approx} once the value of $\tau_\lambda$ is found from the SED fitting procedure:
\begin{equation}\label{dpr_appr}
\dot{M}_{\rm dust}=\frac{\tau_\lambda{\rm (approx)}4 a R_{\rm c} v_{\rm exp} \bar{\rho}_d}{3\bar{Q}_{\rm ext}(\lambda, a)}.
\end{equation}
The condensation radius R$_{\rm c}$ is usually consistently computed by assuming a certain dust temperature at the boundary of the inner shell. 

In the grid of models by \citet{Groenewegen07} the value of T$_{\rm inn}$ is set to be $1000$~K. 
In the Grid of RSG and AGB ModelS (GRAMS) by \citet{Srinivasan11},
the value of R$_{\rm c}$ is independent of T$_{\rm inn}$ and is derived from the fitting procedure.
The grain size are assumed to be all of the same size $a\approx 0.1$~$\mu$m by \citet{Groenewegen07}.
On the other hand, \citet{Srinivasan16} assumed the grain size distribution by \citet{Kim94}.
The average size of the distribution by \citet{Srinivasan16} is around $a\approx 0.1$~$\mu$m. The value of $\bar{Q}_{\rm ext}(\lambda, a)$ is of course both dependent on the optical data set adopted and by the grain size.

The value of $v_{\rm exp}$ is either assumed to be $v_{\rm exp}=10$~km~s$^{-1}$ as in \citet{Groenewegen09} or scaled with the stellar luminosity and the gas-to-dust ratio, $\Psi_{\rm dust}$:
\begin{equation}\label{vexp_scaled}
v_{\rm exp}\propto \left(\frac{L}{L_\odot}\right)^{\alpha} \left(\frac{\Psi_{\rm dust}}{200}\right)^{-\beta},
\end{equation}
where the quantity $\Psi_{\rm dust}$ can be either set equal to 200 for C-rich stars in the SMC \citep{Groenewegen06, Groenewegen07, Groenewegen09, Gullieuszik12, Boyer12, Srinivasan16} or is determined employing some scaling relations \citep{Bressan98, Marigo08}. The power $\alpha$ and $\beta$ can be theoretically derived \citep{Habing94} or empirically determined \citep{vanLoon06_2, Goldman17}.

In the works by \citet{Boyer12} and \citet{Srinivasan16} the value of the expansion velocity is assumed to be $v_{\rm exp}=10$~km~s$^{−1}$ for a star with luminosity $L= 30000$~$L_\odot$ and $\Psi_{\rm dust}=200$. The final DPR for each star is scaled according to Eq.~\ref{vexp_scaled}. 

Once the quantity $\dot{M}_{\rm dust}$ is evaluated through Eq.~\ref{dpr_appr} the mass-loss rate is estimated by assuming a certain value of the gas-to-dust ratio.

Our dust formation scheme allows us to avoid some assumptions adopted in the previous calculations of dusty spectra models and grids.
The most important difference with respect to the previous approach, relies on the consistent evaluation of $\tau_\lambda$, $\Psi_{\rm dust}$, $v_{\rm exp}$. In particular, the optical depth at a given wavelength, condensation radius, dust temperature at the boundary of the dust condensation zone, dust-to-gas ratios and dust density and velocity profiles are computed by our code.   

\subsection{Sample of C-rich stars}
Unless specific filters are used to probe the presence of either O- or C-bearing molecules \citep[e.g.][]{Palmer82, Boyer13}, the identification of C-rich stars in photometric samples is not straightforward.  Fortunately, in the case of the Magellanic Clouds most of the C-rich stars are known to stand out in a particular region of the \ks~vs.~\jks\ color-magnitude diagram (CMD); they are simply indicated as ``C-stars'' in \citet{Boyer11}. These stars are only mildly reddened by the dust present in their CSEs.
Stars more dust-enshrouded are instead selected on the base of their MIR colors and are classified as ``extreme'' (X-) stars.
For 81 stars the spectra from Spitzer's Infrared Spectrograph (IRS) are available \citep{Ruffle15}.

In addition, the catalogue by \citet{Boyer11} contains a distinct class of objects, called ``anomalous'' AGBs (aAGBs) classified on the base of their position in the [8] vs $J-[8]$ diagram. 
Optical medium-resolution spectra were obtained by \citet{Boyer15} using the AAOmega/2dF multi-object spectrograph \citep{Lewis02, Saunders04, Sharp06} for 273 sources, which included 246 aAGBs.
On the base of the analysis performed on the spectra of aAGBs, \citet{Boyer15} concluded that nearly half of the aAGB sample is expected to consist of C-rich stars.
We exclude from the present analysis the sample of aAGBs classified as O-rich and S on the base of the spectral classification, we include instead those for which the classification is uncertain.

We exclude from the SED fitting five sources classified as C-rich on the base of the photometry but not on the base of their IRS or optical spectra.
On the other hand, we include in our analysis five sources photometrically classified as O-rich but as C-rich on the base of their optical spectra plus one source for which the spectral classification is uncertain. The DPRs of these latter sources are included in the C-stars sample.

The catalog by \citet{Boyer11} includes 360 sources classified as FIR sources. This list has been cleaned by \citet{Srinivasan16} on the basis of the IRS classification when available and taking into account the identifications of such sources from other studies. The remaining 33 FIR sources have been divided into 7 groups on the basis of their SED shape and the information on optical/IR variability (see Section 2.4.1 in \citealt{Srinivasan16}). The FIR sources considered as evolved star candidates and included in the dust budget by \citet{Srinivasan16} belong to groups 1$\div$4 (17 sources). Here we exclude the 6 FIR sources spectroscopically classified as O-rich by \citet{Ruffle15}.

The number counts of C-rich sources included in our SED fitting are shown in  Table~\ref{stars_cat} by color class.

In the catalog by \citet{Srinivasan16} the 2MASS NIR photometry is available together with the MIR IRAC, MIPS 24 $\mu$m, AKARI (S11 and L15 filters) survey of the SMC \citep{Ita10}, the WISE (W3 filter) All-Sky data release, the Magellanic Clouds Photometric Survey \citep[MCPS;][]{Zaritsky02} for the $U$ and $B$ bands and the Optical Gravitational Lensing Experiment (OGLE) survey for $V$ and $I$ bands, plus the variability information \citep[OGLE-III;][]{Udalski08}. 

For the NIR bands, the 2MASS photometry of the SAGE-SMC list is matched to data from the InfraRed Survey Facility \citep[IRSF;][]{Kato07} when the photometry is available. 

For the MIR bands (IRAC and MIPS 24 $\mu$m bands), the two epochs of SAGE-SMC have been matched with S$^3$MC epoch for each source. 

\begin{table}
\begin{center}
\caption{Number of fitted AGB stars taken from the catalog by \citet{Srinivasan16}, listed for the different classes of stars, photometrically classified by \citet{Boyer11}. A further selection of the sources is based on their IRS or optical spectra, when available (see text).}
\label{stars_cat}
\begin{tabular}{l c}
\hline
Photometric classification  & Number \\
\hline
C-stars & 1709 \\
X-stars & 339\\
O-stars (C-spectrum) & 5\\
O-stars (unknown spectrum) & 1\\
aAGB  &  1092 \\
FIR &  11 \\
\hline
\end{tabular}
\end{center}
\end{table}

\subsection{SED fitting procedure}
From each of our six grids of spectra we select the best fit model of the sources in Table~\ref{stars_cat} by computing the reduced $\chi^2$ between the modeled and observed photometry, similarly to \citet{Groenewegen09}, \citet{Gullieuszik12}, \citet{Riebel12}, and \citet{Srinivasan16}:
\begin{equation}\label{chi2}
\chi^2=\frac{1}{N_{\rm obs}}\sum_{i}\frac{(F_{i, {\rm obs}} - F_{i, {\rm th}})^2}{e_{i,{\rm obs}}^2},
\end{equation}
where $F_{i, {\rm obs}}$ and $F_{i, {\rm th}}$ are the observed and predicted fluxes for the $i$ band, $e_{i, {\rm obs}}$ is the error of the observed flux and $N_{\rm obs}$ is the number of observed photometric points. The distance of the SMC assumed to compute $F_{i, {\rm th}}$ is $d_{\rm SMC}\approx 60$ kpc \citep{deGrijs15}.
The minimum value of $\chi^2$, corresponding to the best fitting model, is indicated as $\chi^2_{\rm best}$.

For a few sources it is necessary to exclude some of the observed photometric points from the SED fitting procedure. The criteria for the photometry selection is described in the Appendix.

To evaluate the uncertainties of the quantities derived from the SED fitting procedure, we need to estimate the possible variations of the modeled fluxes due to the photometric errors and to the intrinsic stellar variability. While the photometric errors can usually be described by Gaussian distributions, for large-amplitude variables, with observations spanning longer than a period, the distribution of photometric points will have peaks at the extremes of the measured distribution -- just as those expected from sinusoidal-like variations. It is however impossible to identify the dominant factor in determining the error distribution, without performing a star-by-star analysis of the likely periods, amplitudes, and the timespan of the observations in each passband. Therefore, as a compromise between the expected Gaussian and/or sinusoidal distributions, we decide to simply adopt a flat distribution of errors, distributed over the magnitude interval of the observations. This conservative assumption probably provides an upper limit to the actual error distribution. 

We therefore generate 100 sets of $N_{\rm ph}$ random numbers, $r_{\rm i}$, where $N_{\rm ph}$ is the number of photometric bands, extracted from a uniform distribution of values between $-1$ and $1$.
For each star, we only consider the valid photometric points, $i$, and we add to our best fit spectrum the quantity $r_i\times e_{\rm i, {\rm obs}}$:
\begin{equation}\label{error}
    F_{\rm i,{\rm th+err}}=F_{\rm i, {\rm th}}+ r_{\rm i} \times e_{\rm i, {\rm obs}}.
\end{equation}
For all the 100 sets of random generated numbers we then re-calculate the $\chi^2$ of $F_{i, {\rm th+err}}$ through Eq.~\ref{chi2}.
We finally extract the minimum and the 34$^{\rm th}$ $\chi^2$ values corresponding to 1$\sigma$ variation, which provides an estimate of variation of the $\chi^2$ within 1$\sigma$ ($\Delta\chi^2$).

We compute the quantity $\chi^2_{\rm max}=\chi^2_{\rm best}+\Delta\chi^2$ and we extract, for each star, the models in the grid with $\chi^2 \le \chi^2_{\rm max}$. We then compute the
average value of each of the quantities for the valid models, i.e. expansion velocity, gas-to-dust ratio, as well as the input stellar quantities, i.e. luminosity and effective temperature. 
For estimating the uncertainties over each of the quantities, we calculate the dispersion $\sigma$ from the average value.
If the number of models with  $\chi^2 \le \chi^2_{\rm max}$ is $\le 3$, we assume as average value the quantity corresponding to the $\chi^2_{\rm best}$ while we set the uncertainty equal to zero.

\section{Results}
\label{Results}
\subsection{Trends in the grids of dusty models}\label{trends_tau}
We here discuss some trends between the optical depth at $\lambda=1$ $\mu$m, $\tau_1$, where the spectra of TP-AGB stars peak, and the main stellar parameters such as $L$, carbon-excess, $M$,  $T_{\rm eff}$, and mass-loss rate.
We opt to show $\tau_1$ because, as discussed in other works \citep[e.g., ][]{Bressan98, Nanni16}, the SED of a dust enshrouded star is mainly shaped by this quantity.
From Eqs.~\ref{tau_lambda} and \ref{tau_approx} we can see that $\tau_1$ is dependent on several quantities, such as $\delta_{\rm i}$, $R_{\rm c}$ and $v_{\rm exp}$ which, in turn, are correlated with the stellar parameters.

\begin{figure}
\includegraphics[trim=0 0 0 0]{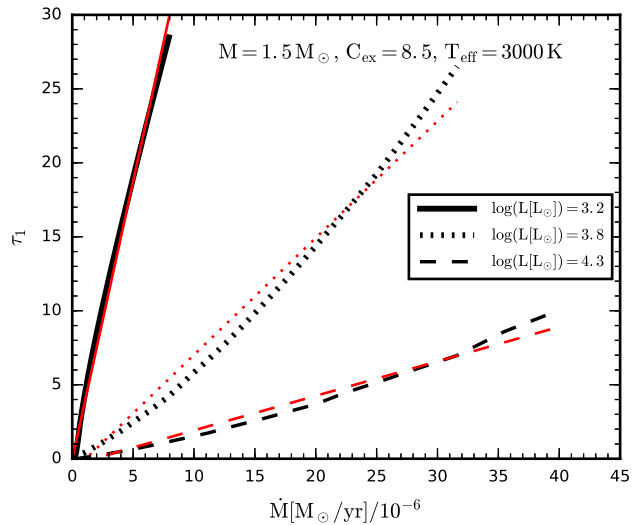}
        \caption{Optical depth at 1~$\mu$m, $\tau_1$, as a function of mass-loss rate, for different choices of the luminosity, listed in the legend. Black thick lines are the trends derived from our models, whilst the corresponding red thin lines are the linear fit between $\tau_1$ and $\dot{M}$. The models shown are computed with R12 optical data set for carbon dust. The other stellar parameters of the selected models are mentioned in the figure.}
        \label{tau1_dmdt}
        \end{figure}

In Fig.~\ref{tau1_dmdt} we plot $\tau_1$ against $\dot{M}$ for different $L$ computed for R12 optical data set. Thick black lines represent the trends derived from our models, whilst the thin red lines are the linear fits between $\tau_1$ and $\dot{M}$. %leo changed: $\dot{M}/10^{-6}$.
The mass-loss has large impact on the final optical depth, as can be also seen from Eqs.~\ref{tau_lambda} and \ref{tau_approx}.
However, the dependence of $\tau_1$ on the mass-loss rate depends on the value of the luminosity, with a flatter trend for larger luminosity.
Such a trend depends on the fact that the fraction of the condensed dust, which determines $\delta_{\rm i}$ in Eq.~\ref{tau_lambda}, is not constant in our models but changes as a function of the input stellar parameters. 

\begin{figure}
\includegraphics[trim=0 0 0 0]{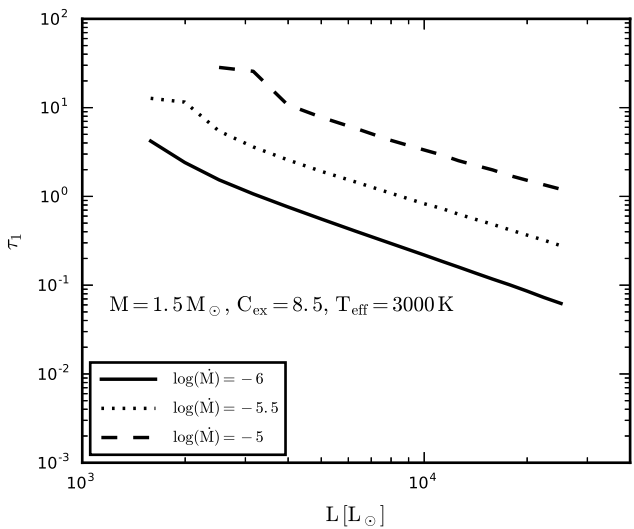}
        \caption{Optical depth at 1~$\mu$m, $\tau_1$, as a function of luminosity, for different choices of the mass-loss rate, listed in the legend. The models shown are computed with R12 optical data set for carbon. The other stellar parameters of the selected models are mentioned in the figure.}
        \label{tau1_L}
        \end{figure}

\begin{figure}
\includegraphics[trim=0 0 0 0]{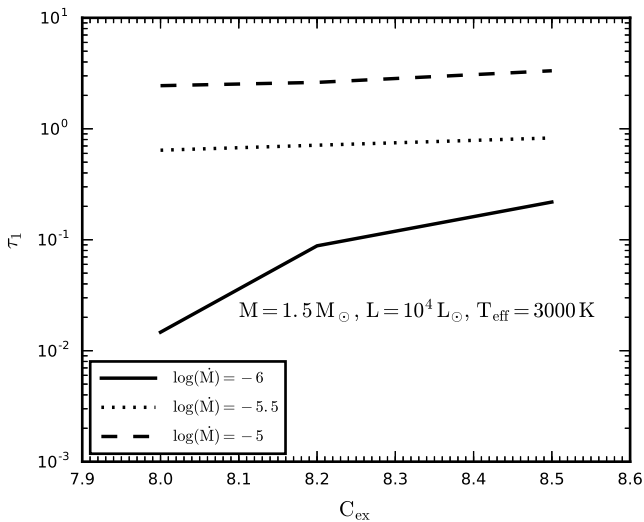}
        \caption{The same as Fig.~\ref{tau1_L} but with $\tau_1$, as a function of carbon-excess.}
        \label{tau1_CO}
        \end{figure}

In Figs.~\ref{tau1_L} and \ref{tau1_CO}, $\tau_1$ is shown as a function of the luminosity and carbon-excess, respectively, for different choices of the mass-loss rate. In these two figures, we show models computed for the R12 optical data set.

We can clearly see from Figs.~\ref{tau1_L} and \ref{tau1_CO}, that, for a given value of $\dot{M}$, the quantity $\tau_1$ is mostly changed by the stellar luminosity, while the carbon excess usually produces only a secondary effect in the range of values considered, except for the smallest mass-loss rate. 
The strong dependence of $\tau_1$ on the stellar luminosity is qualitatively in agreement with the scaling relations found by \citet{Bressan98} and \citet{Ivezic10}.
We note, however, that the dependence of $\tau_1$ on the quantity $L$ is slightly stronger than the one predicted by \citet{Bressan98,Ivezic10}. Assuming a dependence $\tau_1\propto L^{-\alpha}$ we obtain $\alpha=-1.45, -1.22, -1.16$ for $\log\dot{M}=-6, -5.5,-5$, respectively, rather than $\alpha=-0.85$ of \citet{Bressan98} or $\alpha=-1$ of \citet{Ivezic10}.
The dependence of $\tau_1$ as a function of $L$ is related to the behaviour of $R_{\rm c}$, $v_{\rm exp}$ and $\delta_{\rm i}$ as a function of $L$ and for different mass-loss rates (Eqs.~\ref{tau_lambda} and \ref{tau_approx}).
In particular, by inspecting the models shown in the figures, we find that R$_{\rm c}$ increases more steeply with $L$ for the lowest mass-loss rate than for the larger ones, whilst the condensation fraction decreases more rapidly for the lowest mass-loss rate plotted.
On the other hand, the outflow is not accelerated for $\log\dot{M}=-6$ so that the velocity does not depend on the luminosity for this specific case. 

For the largest mass-loss rates considered, the velocity increases as a function of the luminosity and the condensation of dust is partially inhibited by the outflow acceleration, which dilutes the gas. 
However, the gas density is always large enough that the amount of dust condensed decreases less rapidly as a function of $L$ than for the low mass-loss case.
All these trends between the different stellar quantities and $L$ explain the steeper dependence recovered between $\tau_1$ and $L$ for the lowest mass-loss rates with respect to the larger ones.

We always find a mild dependence of $\tau_1$ on $M$, with $\tau_1$ only slightly increasing as a function of the stellar mass.
Finally, the effect of T$_{\rm eff}$ on $\tau_1$ is also quite mild, even though for the largest mass-loss rates $\tau_1$ tends to decrease for increasing T$_{\rm eff}$.

\subsection{Quality of the fit}
The six grids of models produced are able to provide a good SED fit for most of the sources.
In order to guarantee a good estimate of the total DPR, we analyze the $\chi^2_{\rm best}$ of the most dust producing stars among the C- and X-stars, yielding $\approx80\%$ of the total DPR. The procedure is similar to the one applied by \citet{Srinivasan16}.
The $\chi^2_{\rm best}$ of these sources is $\lessapprox 1$ for the $\approx42-52\%$ of the sample considered and $\lessapprox 10$ in the $\approx94-98\%$ of the cases.
Only one among the most dust-enshrouded stars (IRAS 00350-7436) is very poorly fitted by all the optical data sets of carbon dust ($\chi^2_{\rm best}>100$). 
This source has also been suggested to be a Post-AGB star by \citet{Matsuura05} and
was discussed by \citet{Srinivasan16}, who also were not able to provide a good fit.

\begin{figure}
\includegraphics[trim=0 0 0 -2cm, angle=90, width=0.48\textwidth]{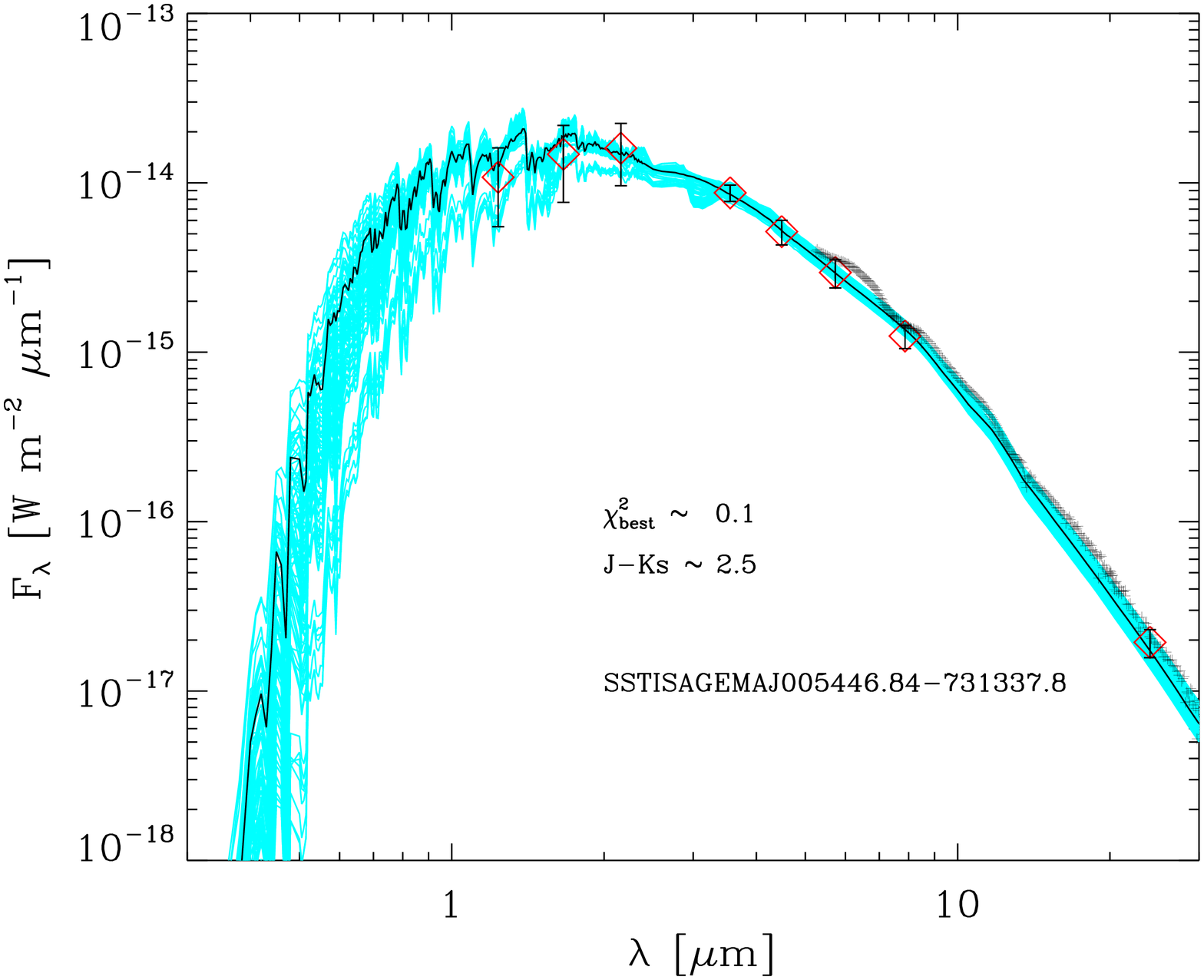}
\includegraphics[trim=0 0 0 -2cm, angle=90, width=0.48\textwidth]{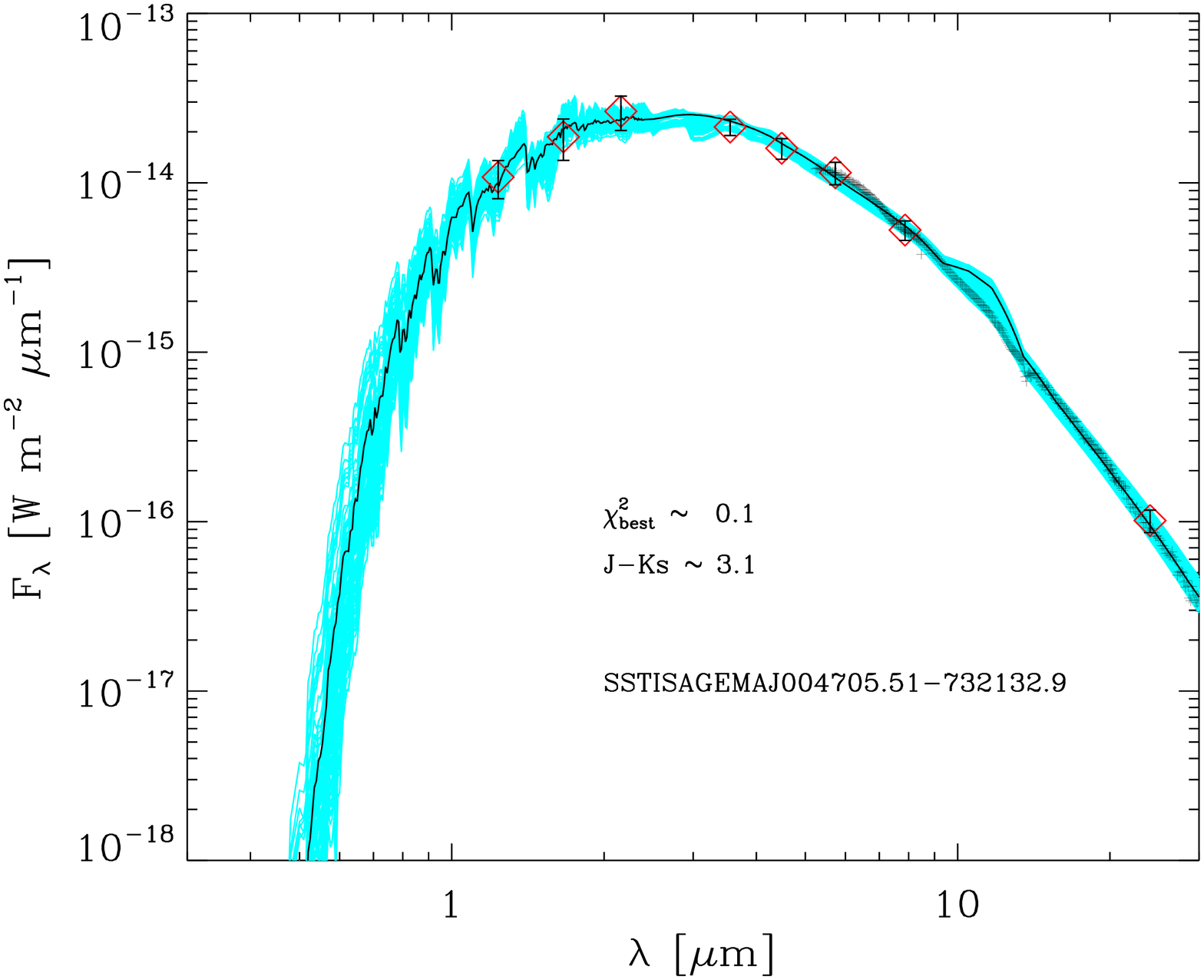}
        \caption{Two examples of carbon-rich stars well fitted with R12 data set for carbon dust. Red diamonds represent the valid observed photometric data points, overplotted with their uncertainties (error bars). The black solid lines represent the best-fitting spectra. The spectra overplotted in cyan correspond to models producing an acceptable $\chi^2$, as explained in the text. The values of \jks\ and $\chi^2_{\rm best}$ for each star are written in the figures. The IRS spectrum of each star is also overplotted with black crosses. }
        \label{good_fit}
        \end{figure}

\begin{figure}
\includegraphics[trim=0 0 0 -2cm, angle=90, width=0.48\textwidth]{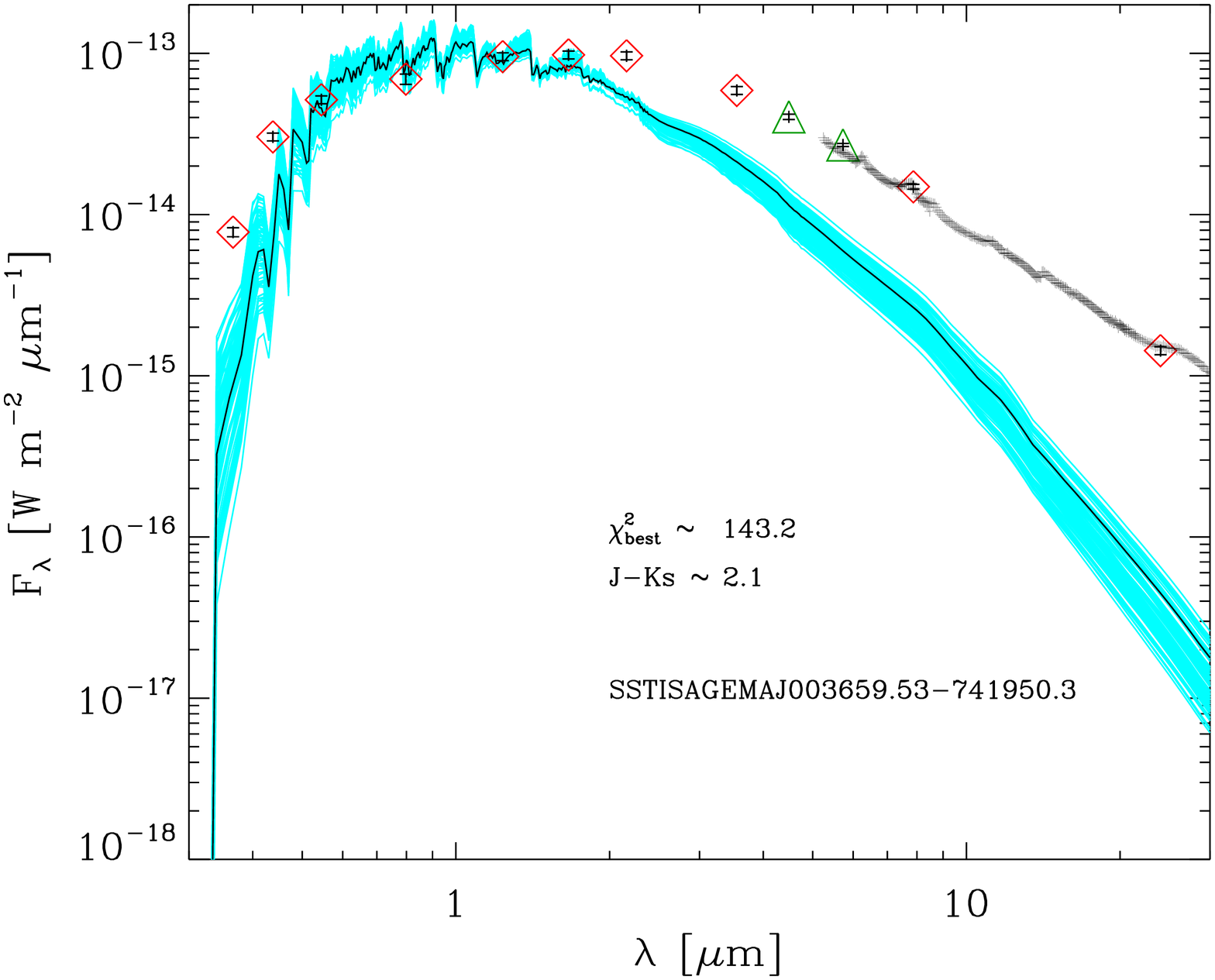}
\includegraphics[trim=0 0 0 -2cm, angle=90, width=0.48\textwidth]{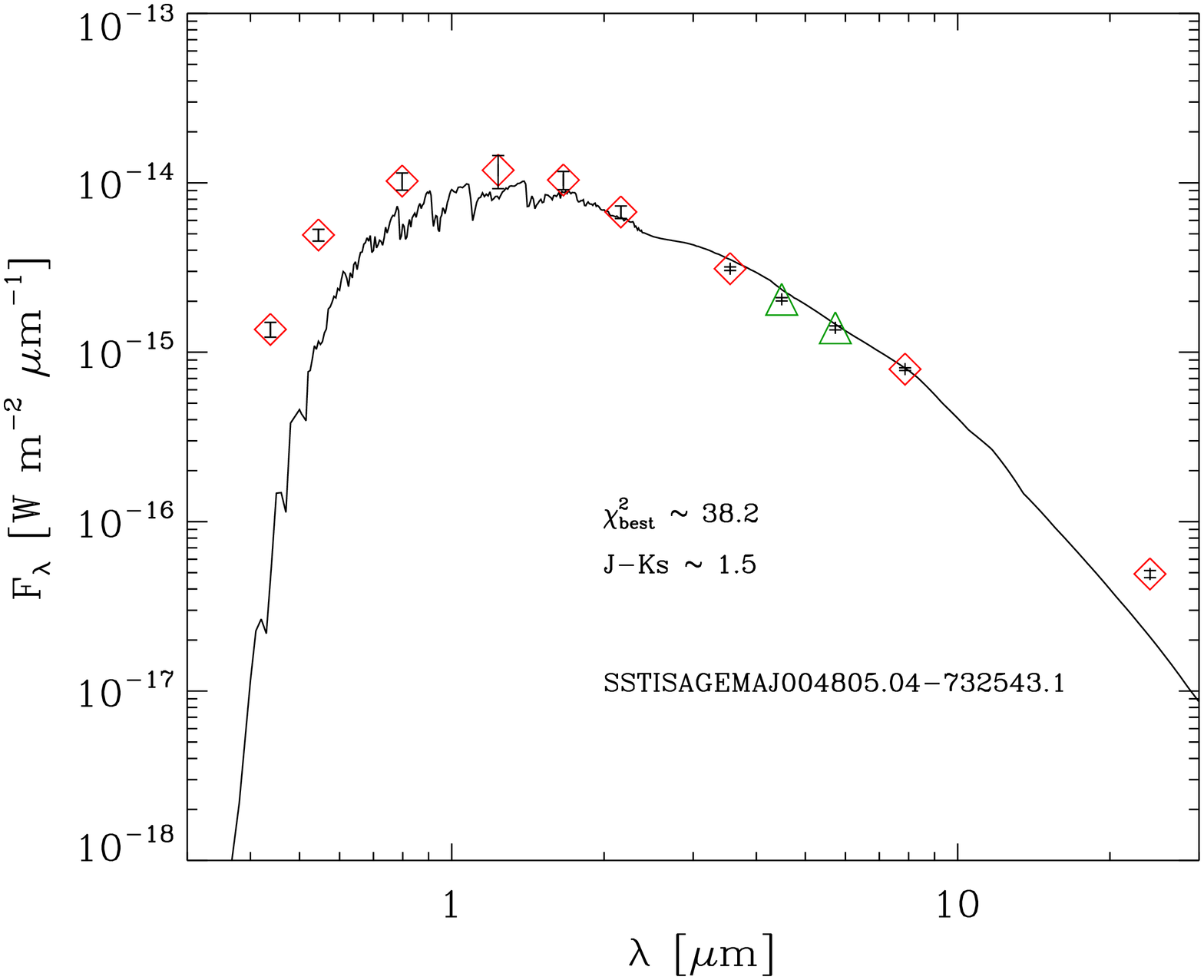}
        \caption{The same as Fig.~\ref{good_fit}, but for two carbon-rich stars not well fitted by the theoretical spectra. Red diamonds represent the valid observed photometric data points over plotted with their uncertainties, whilst the two green triangles are the photometric points excluded from the fit on the base of the criteria discussed in the Appendix. The source in the upper panel is IRAS 0035-7436, discussed in the text. No IRS spectrum is available for the source in the lower panel. Only one valid theoretical spectrum is derived from the analysis, drawn in solid black.}
\label{bad_fit}
\end{figure}

In Figs.~\ref{good_fit} and~\ref{bad_fit} we show two examples of well fitted C-rich stars, with $\chi^2_{\rm best}\approx0.1$, and two examples of stars with a poor fit, respectively. The valid observed photometric data points and uncertainties are plotted with red diamonds, whilst the photometric points excluded with the criteria described in the Appendix, are plotted with green triangles. The IRS spectrum of each source has been overplotted when available (black crosses). The solid black line is the theoretical spectrum from our grid of models corresponding to the best fit. The cyan lines are the spectra of models producing a $\chi^2\le \chi^2_{\rm best}+\Delta\chi^2$.
In the upper panel of Fig.~\ref{bad_fit} the star previously discussed, IRAS 00350-7436, is shown.

For the aAGBs sample $\chi^2_{\rm best}$ is $\lessapprox 1$ and  $\lessapprox 10$ for about the $\approx20\%$ and $\approx95\%$ of the cases, respectively.
Even though \citet{Boyer15} argued that half of these stars are possible oxygen-rich we are able to provide an acceptable fit for all of them employing our carbon-rich models.

On the other hand, the fit is not always that satisfactory for the FIR, with $\chi^2_{\rm best}\gtrapprox10$ for five sources. 

\subsection{Dust properties and stellar quantities}
From the SED fitting procedure we can derive some important stellar quantities and dust properties.
\subsubsection{Luminosity function}
\begin{figure}
\includegraphics[
width=0.45\textwidth]{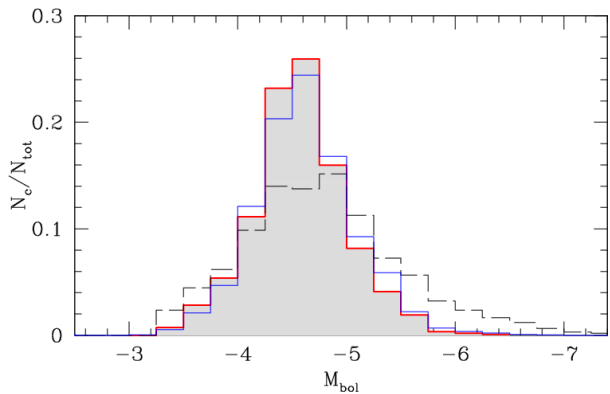}
        \caption{Luminosity function of C- and X-stars
derived from the SED fitting procedure (red histogram shaded in grey). For comparison we plot the normalized distributions of C-stars obtained from the catalog of \citet{Rebeirot_etal93}
(dashed histogram), and that derived by \citet{Srinivasan16} using their best-fit luminosity estimates (blue histogram).}
        \label{L_func}
        \end{figure}

In Fig.~\ref{L_func} the luminosity function of the C- and X-stars is shown. The optical data set selected is R12, but the result does not change significantly with a different data set.
The absolute bolometric magnitude ranges between M$_{\rm bol}\approx-3.2$ and M$_{\rm bol}\approx-6.2$ while the peak of the distribution is around M$_{\rm bol}\approx-4.5$.
Our distribution is in excellent agreement with the one
obtained by \citet{Srinivasan16} through SED fitting applied to same sample of stars. For comparison we also show the luminosity function of C-stars in the SMC obtained from the catalog of \citet{Rebeirot_etal93}, which includes 1707 C-stars identified on GRISM plates. We obtain the bolometric magnitudes from the visual magnitudes at 5220~\AA\ using the corrections by \citet{Westerlund_etal86} and assuming a true distance modulus for the SMC $\mu_0 = 18.9$~mag. For consistency of comparison we exclude the faintest (M$_{\rm bol} > -3.2$) C-stars from the  \citet{Rebeirot_etal93} catalog since our sample includes C-star brighter than the RGB-tip.
We note that the peak location of the luminosity function of \citet{Rebeirot_etal93} agrees with ours and with \citet{Srinivasan16} at  M$_{\rm bol} \sim -4.5$, but the distribution is broader, extending towards fainter and brighter magnitudes. The reason for this difference is not clear, likely related to the different methods.

\subsubsection[trim= 1cm 0 0 0]{Mass-loss rates}\label{sec:ml}
\begin{figure*}
\includegraphics[trim= 1cm 0 0 0]{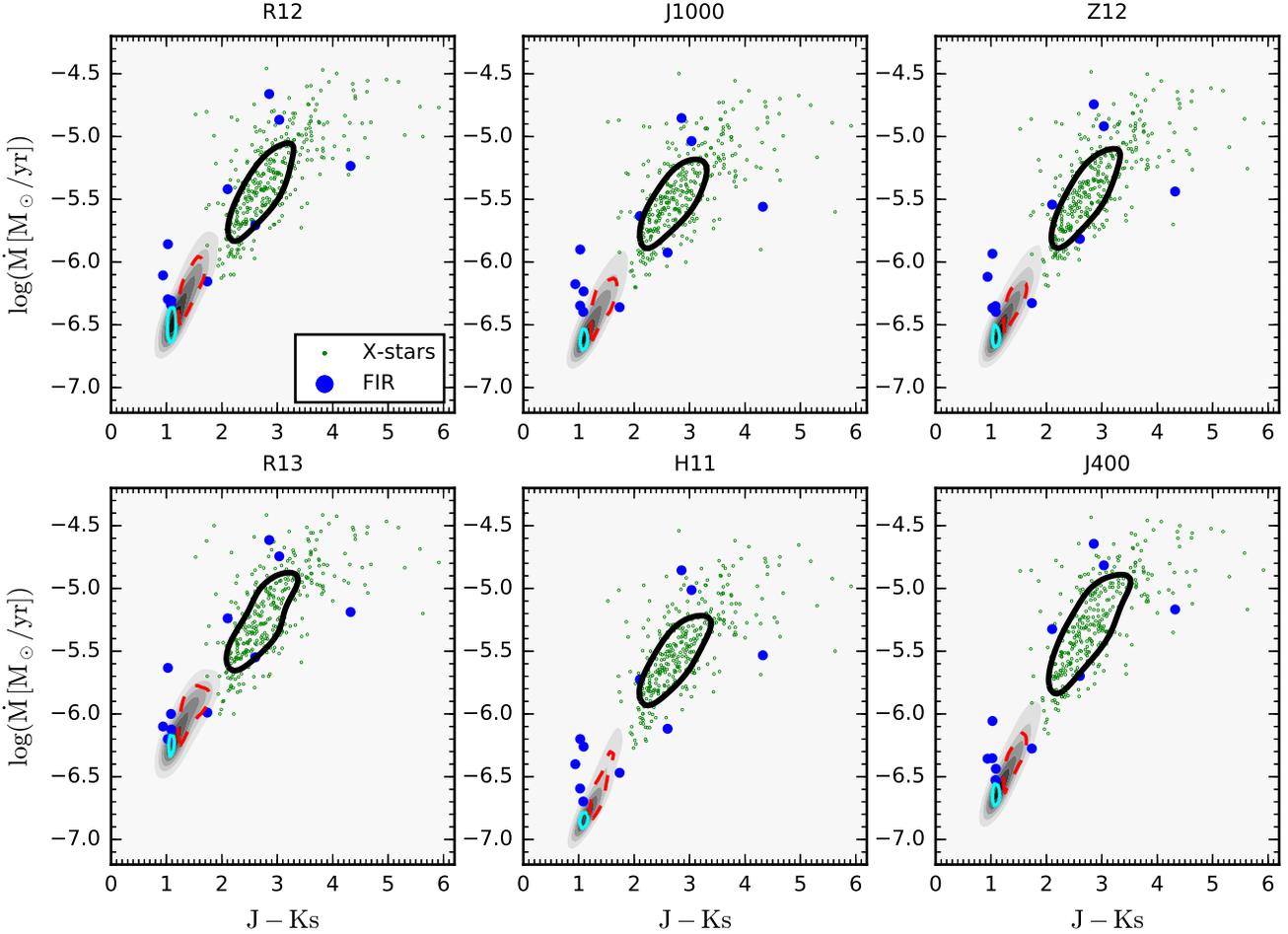}
        \caption{Mass-loss rate as a function of the \jks\ color derived from the SED fitting procedure for the different optical data sets. The linear normalized density map from 0, light grey, to 1, black, includes all classes of stars. X- and FIR stars are over plotted with different colors and symbols listed in the legend. C-, X- and aAGB stars are also contour plotted with different colors and line styles: dashed red for C-stars, solid black thick for X-stars and solid cyan thin for aAGBs.}
        \label{ml}
        \end{figure*}

\begin{figure}
\includegraphics[trim=0 -0.5cm 0 0]{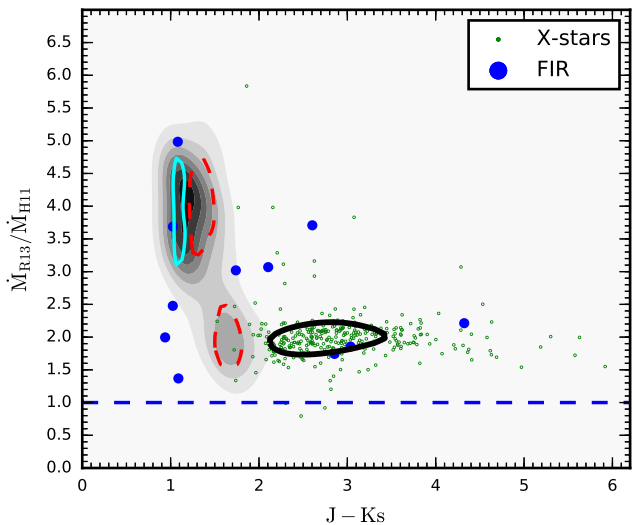}
        \caption{Ratios between the mass-loss rates derived for R13 and H11 data sets as a function of \jks\ color. The color code for the different classes of stars is the same as in Fig. \ref{ml}. The blue dashed line indicates a constant ratio, $=1$, corresponding to the same mass-loss rate for the two data sets.}
        \label{ml_op}
        \end{figure}

\begin{figure*}
\includegraphics[trim= 1cm 0 0 0]{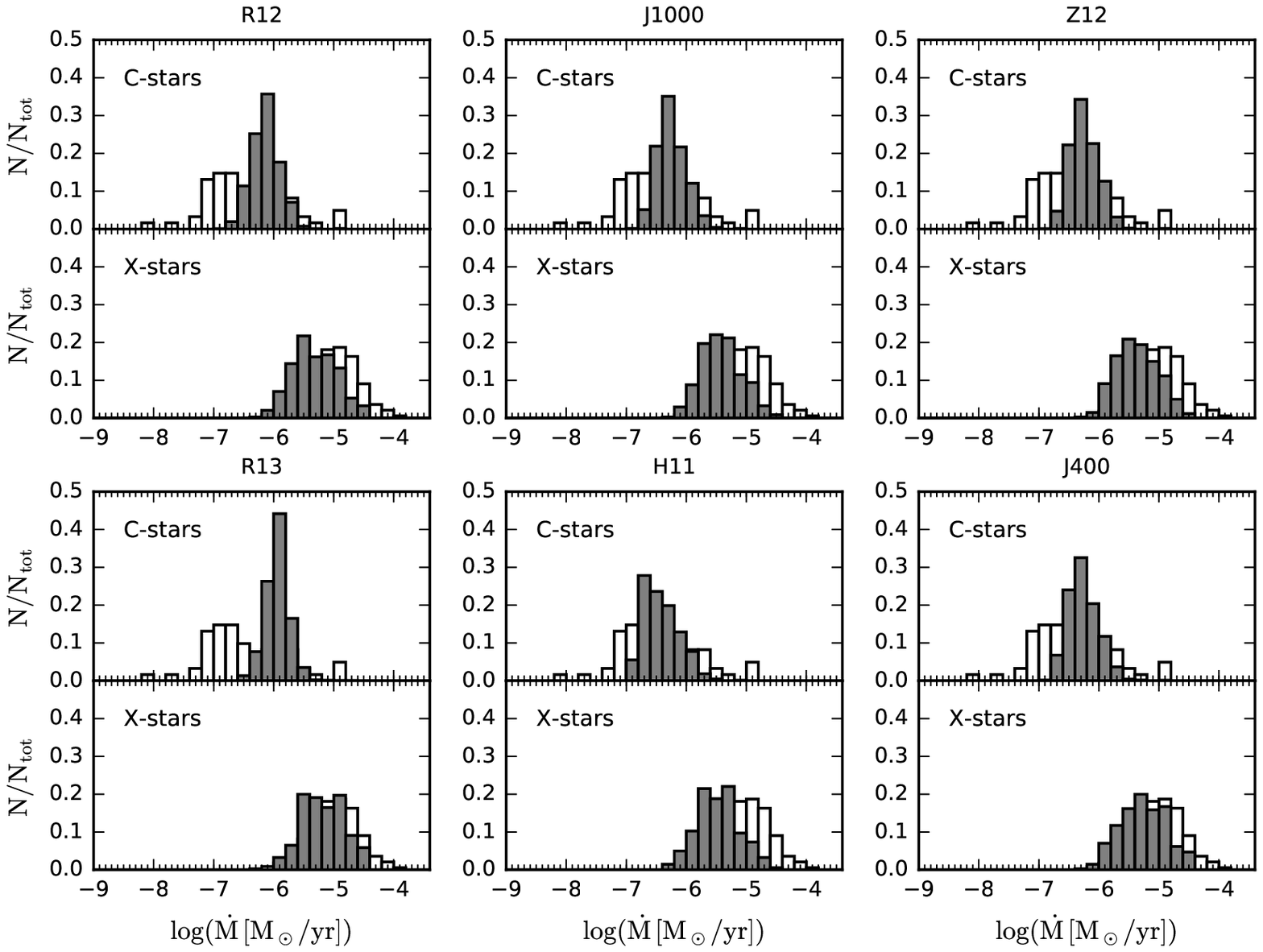}
        \caption{For each chart, upper panel, the normalized distribution of mass-loss rates for the stars in the SMC (shaded histogram) are compared to that of a volume-limited sample of Galactic optically bright C-stars from \citet{Schoier01} (white histogram). In the lower panel of each chart, X-stars in the SMC are compared to a sample of Galactic highly-obscured C-stars \citep{Groenewegen02}.}
        \label{ml_comp}
        \end{figure*}

In the six panels of Fig.~\ref{ml} the linear, normalized density maps of all the sources are plotted against the \jks\ color. The different classes of stars are contour lined with different styles.

The mass-loss rates derived for aAGBs are typically lower than the ones of C-stars. 
The FIR sources are not located in a specific place in the diagram and their position depends on \jks\ .
The separation between C- and X-stars occurs around $\log\dot{M}\approx-6$ at $\jks\approx 2$. Stars with $\jks\gtrapprox3$ reach the largest mass-loss rates of $-4.6\lessapprox\log\dot{M}\lessapprox-4.4$, where the exact values depend on the optical data set considered. 

The mass-loss rates for the different classes of stars with R13 and H11 are respectively the largest and the lowest derived by our analysis.
With R13 we obtain $\log\dot{M}\approx-6.3$, for aAGBs, $-6.2\lessapprox\log\dot{M}\lessapprox-5.8$, for C-stars, and $-5.6\lessapprox\log\dot{M}\lessapprox-4.8$, for X-stars.
For H11, aAGBs are located around $\log\dot{M}\approx-6.8$, C-stars between $-6.8\lessapprox\log\dot{M}\lessapprox-6.3$ and X-stars between $-5.9\lessapprox\log\dot{M}\lessapprox-5.2$.
For R12, J1000, Z12 and J400 each class of stars occupies approximately the same region in the plot. For these data sets, aAGBs are concentrated around $\log\dot{M}\lessapprox-6.4$, whilst most of the C- and X-stars are located between $-6.5\lessapprox\log\dot{M}\lessapprox-6$ and $-5.8\lessapprox\log\dot{M}\lessapprox-5$, respectively.

The uncertainty affecting the mass-loss estimate is between $\approx$10 and $\approx$60$\%$ for the least dust-rich stars and between $\approx$20 and $\approx$40$\%$ for dust enshrouded stars.

In Fig.~\ref{ml_op} we plot the ratio of the mass-loss rates derived for R13 and H11, as a function of the \jks. As discussed before, these two data sets yield the maximum and minimum typical values of the mass-loss rates.
The figure points out that the mass-loss rates of the least dusty stars are typically between $\approx1.5-5$ times larger for R13, whilst, for the dustiest stars, the mass-loss rate is typically a factor of $\approx2$ larger for R13. 
For few C-stars, not visible in the density plot, the mass-loss rates derived with R13 can be up to $\approx$7-8 times larger than the ones obtained with H11.
Larger mass-losses for the R13 data set, implies that a higher density than for H11 is needed to produce approximately the same dust extinction.

In Fig.~\ref{ml_comp} the normalized distributions of mass-loss rates derived for all the optical data sets are shown for both groups of C- and X-stars. These two classes of stars are compared to those of a Galactic samples taken from the literature. {Dust enshrouded stars are taken from \citet{Groenewegen02}, while optically-bright C-stars are from \citep{Schoier01}.
The most dusty stars in our Galaxy share comparable distributions with X-stars in the SMC, especially for R13 and J400 data sets, even though the peak of the SMC distribution is shifted to lower values of the mass-loss rates.
Optically-bright C-stars in the SMC are characterized by somewhat larger mass-loss rates compared to the Galactic ones, whose distribution is broader and extends down to lower values. The only exception is represented by H11, for which the mass-loss distribution of C-stars in the SMC is similar to the one of optically bright Galactic sources.

\subsubsection{Gas-to-dust ratios}\label{sec:dtg}
We consistently derived the values of the gas-to-dust ratio (Eqs.~\ref{delta_i},~\ref{d_mloss} and~\ref{psi_dust}) and the associated uncertainties for the fitted stars.

In Fig.~\ref{dg} we show the gas-to-dust ratios as a function of the mass-loss rate for all the optical data sets. The value of $\Psi_{\rm dust}$ can differ considerably from the constant value of $200$ usually adopted \citep{Groenewegen06, Groenewegen07, Groenewegen09, Gullieuszik12, Boyer12, Srinivasan16}.
In fact, $\Psi_{\rm dust}$ spans a large range of values, especially for the sources with $\log\dot{M}\lessapprox-6$.
The value of the mass-loss rate $\log(\dot{M})\approx-6$ ($\jks\approx 2$) roughly corresponds to the the separation between C- and X-stars, appearing around $\Psi_{\rm dust}\approx 800-2000$, depending on the optical data set.
For $\log\dot{M}\lessapprox-6$, $\Psi_{\rm dust}$ increases steeply as a function of the mass-loss rate with progressively less dust in CSEs.
Stars with $\log\dot{M}\gtrapprox -6$, form dust more efficiently in their CSEs. This result indicates that dust condensation is more efficient in denser environments, which correspond to larger values of the mass-loss rate. Furthermore, the dependence between $\Psi_{\rm dust}$ and the mass-loss rate is much milder than for $\log\dot{M}\gtrapprox-6$ and $\Psi_{\rm dust}$ tends to saturate.
Most of X-stars have gas-to-dust ratios of $800\lessapprox\Psi_{\rm dust}\lessapprox2000$, depending on the selected opacity.
The quantity $\Psi_{\rm dust}$ for X-stars derived with R12 and J400 is less scattered than the other data sets. The lowest value of $\Psi_{\rm dust}$ also depends on the opacity adopted, but it is usually no lower than $\approx 500$, with only few exceptional stars. In any case, $\Psi_{\rm dust}$ is never as low as the value of 200, usually assumed in the literature.

The predicted values of $\Psi_{\rm dust}$ of aAGBs are much larger than the ones of C- and X-stars, as also expected from the analysis performed by \citet{Boyer11}. Anomalous AGBs are in fact expected to be very inefficient dust producers. On the other hand, FIR sources are not placed in specific regions of the plots.

The uncertainty associated to the gas-to-dust ratio, is larger for the least dusty stars, usually between $\approx20$ and $\approx80\%$, whilst it is between $\approx30$ and $\approx40\%$ for the dustiest sources.
The largest uncertainty in the gas-to-dust ratio for the least dust enshrouded stars is not surprising, since large values of the gas-to-dust ratio are always expected to correspond to almost dust-free spectra, which yield equally good fits for almost dust-free stars.

In Fig.~\ref{gtd_R13_R12} we show the ratio between $\Psi_{\rm dust}$ obtained with R13 and R12. These data sets yield among the most different values of $\Psi_{\rm dust}$, especially for the dustiest stars. For the least dusty stars, the gas-to-dust ratio for R13 is between $\approx1.2-2.8$ times larger than the one obtained with R12. For some sources, not visible in the density plot, $\Psi_{\rm dust}$ is up to $\approx7$ times larger for R13. For X-stars, the $\Psi_{\rm dust}$ of R13 is usually between $\approx40$ and $\approx90\%$ larger than the one of R12.
This result implies that dust condensation is expected to be more efficient if the type of dust formed has optical constant similar to R12 rather than R13.

The spread in the values of $\Psi_{\rm dust}$ predicted by our analysis is also found in the most updated grid of dynamical models by \citet{Eriksson14}, in which the value of $\Psi_{\rm dust}$ can vary by a factor 30, with $330\lessapprox\Psi_{\rm dust}\lessapprox10000$. 
The value of $\Psi_{\rm dust}$ is expected to be related to the carbon excess, with lower $\Psi_{\rm dust}$ for larger values of the carbon-excess \citep{Mattsson10,Nanni13,Eriksson14}.
This result again shows that a unique value of $\Psi_{\rm dust}$ for all the C-rich sources might not be a realistic choice, as also noticed by \citet{Eriksson14}.

        \begin{figure*}
\includegraphics[trim= 1cm 0 0 0]{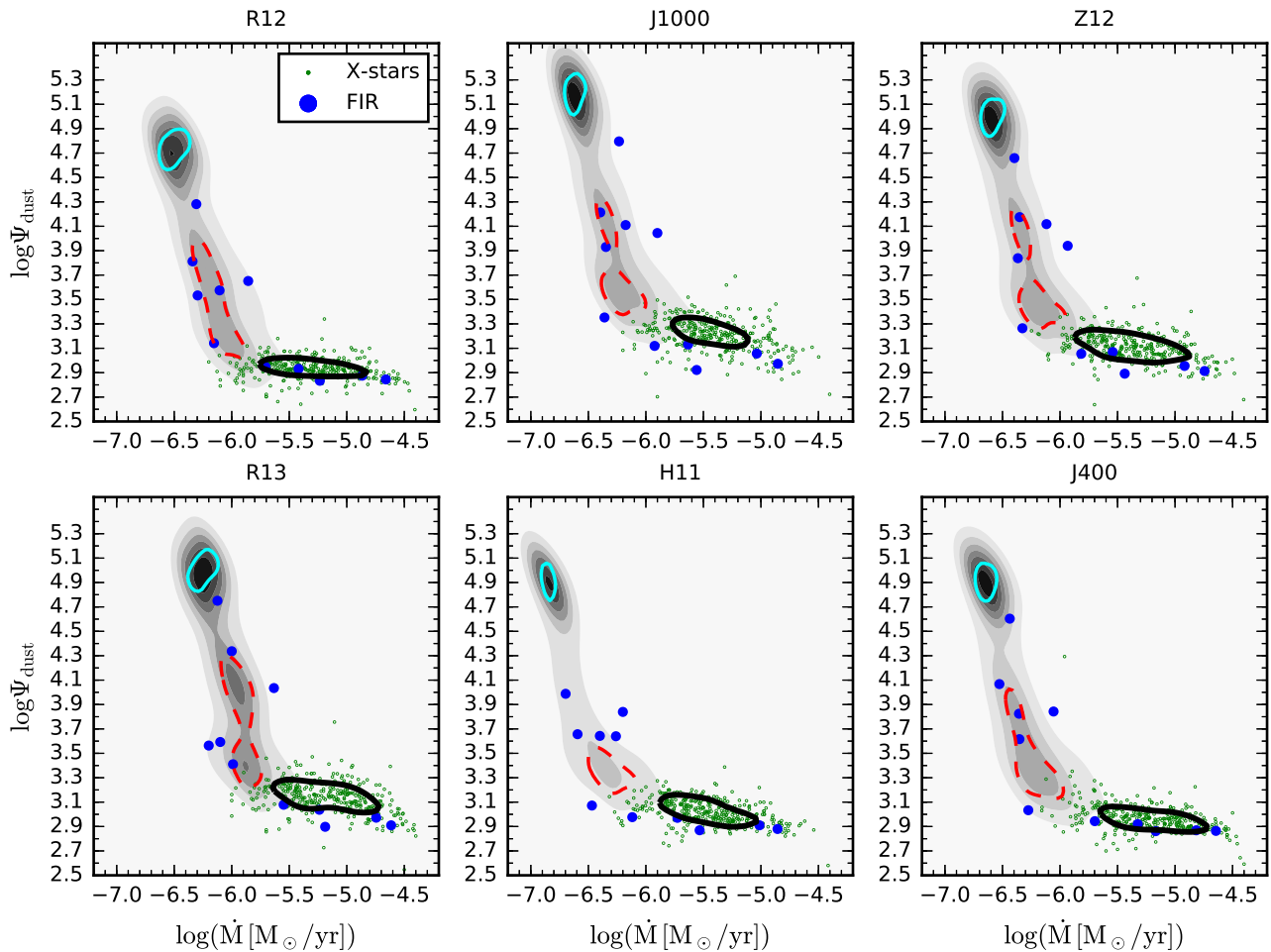}
        \caption{Gas-to-dust ratio as a function of the mass-loss rate for all the optical data sets of carbon dust. The color code for the different classes of stars is the same as Fig.~\ref{ml}.}
        \label{dg}
        \end{figure*}
        
                        \begin{figure}
\includegraphics[trim=0 -0.5cm 0 0]{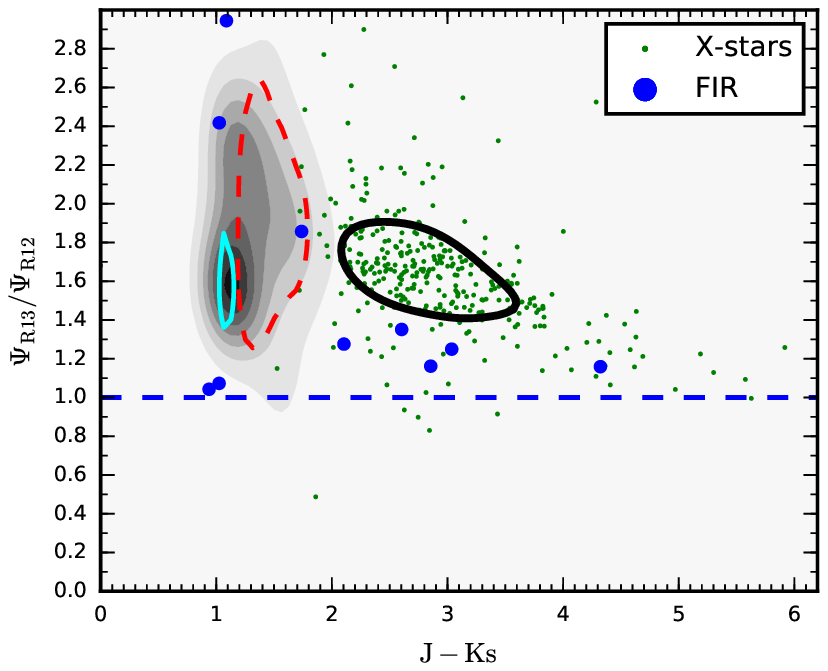}
        \caption{Ratios between the values of $\Psi_{\rm dust}$ derived for R13 and R12 data sets as a function of \jks\ color. The color code for the different classes of stars is the same as in Fig.~\ref{ml}. The blue dashed line indicates a constant ratio, $=1$, corresponding to the same value of $\Psi_{\rm dust}$ rate for the two data sets.}
        \label{gtd_R13_R12}
        \end{figure}
        
\subsubsection{Dust chemistry}
The growth of all the dust species included in our model (amorphous carbon, SiC and metallic iron) is followed with Eq.~\ref{dadt}.
Since the growth rate of a certain dust species is proportional to the number density of the dust-forming molecules in the CSEs, we expect the mass fraction of SiC and iron to be dependent on the starting abundance of Si and Fe atoms, respectively, on the density of the outflow, determined through the mass-loss rate and the velocity (Eq.~\ref{dens}), and on formation sequence of the dust species. The abundances of Si and Fe depend on the metallicity of the SMC. On the other hand, the formation of carbon dust is, in many cases, able to accelerate the outflow and to reduce the density. The condensation sequence of dust in our model, usually predicts SiC to form first, before that the outflow acceleration, followed by amorphous carbon and iron. This latter dust species is formed after that the outflow is accelerated, when this process occurs.
Similarly to the the amorphous carbon dust, the final grain sizes of SiC and iron are also dependent on the seed particle abundance.

In Fig.~\ref{dust_chem} the mass fractions of SiC over the total dust produced is plotted against \jks\ for R12 data set. The trend found is not much dependent on the adopted set of optical constants.
This figure highlights a trend between the SiC mass fraction and \jks\ for X-stars, which saturates around $\jks\approx3$ and never becomes larger than $\approx 10\%$. 
This trend suggests that more SiC is formed in denser environments, since larger values of \jks\ corresponds to larger mass-loss rates. In fact, since SiC forms before carbon dust, this dust component forms before the onset of the dust driven-wind and more condensation is expected for larger densities. On the other hand, the mass-fraction of SiC is limited by the initial abundance of Si, dependent on the metallicity. As a consequence, an upper value of the SiC mass fraction is attained around \jks$\approx3$.
For less dust rich stars, a reversed trend between the SiC mass fraction and \jks\ is visible. However, since the values are always below $\approx 1\%$ the trend is not well defined.
The uncertainty associated to the SiC mass fraction can be large even for dust enshrouded stars. For $\jks\approx3$ the uncertainty is between $\approx 20$ and $\approx 100\%$ and it decreases with $\jks$, down to $\approx 30\%$.

We do not find a clear trend between the iron dust mass fraction with the \jks\ color, and its fraction is always below $1\%$ for all the classes of stars. 
The low amount of iron dust is an expected result, since this dust species is formed later than amorphous carbon when to the outflow density decreases following the accelerated expansion.

\begin{figure}
\includegraphics[trim=0 0 0 0]{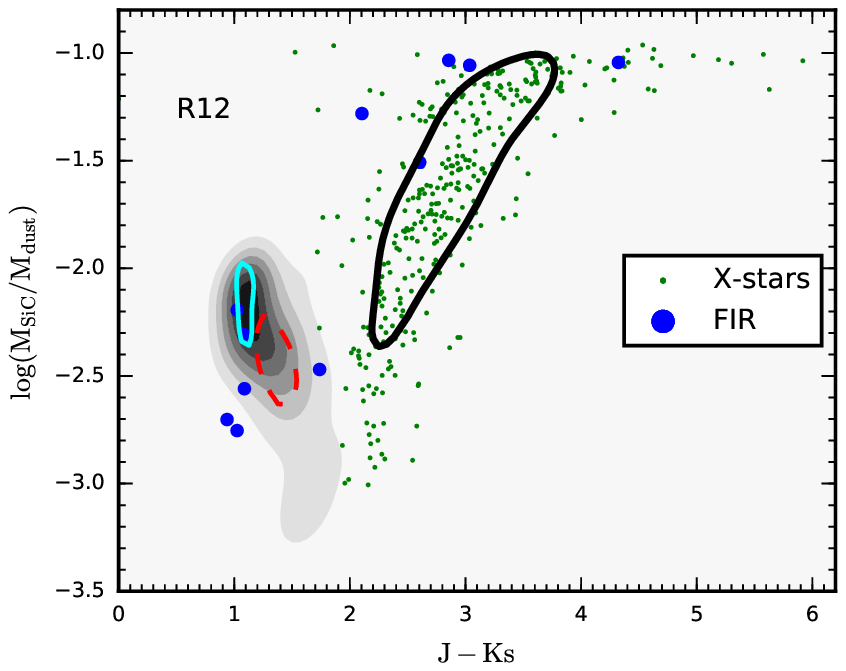}
        \caption{SiC mass fraction as a function of \jks\ derived with R12 optical data set. The color code for the different classes of stars is the same as in Fig.~\ref{ml}.}
        \label{dust_chem}
        \end{figure}

\subsubsection{Grain sizes}\label{sec:grains}
In Fig.~\ref{rc} we plot the size of carbon grains obtained for the sources fitted with R13 against \jks\, for the different classes of stars.
Stars with increasing dust content attain larger grain sizes.
For R13 data set, amC dust grains in CSEs of aAGBs are usually around $0.03\lessapprox a_{\rm amC}\lessapprox0.04$ $\mu$m. On the other hand, the bulk of C- and X-stars are fitted with models with final grain sizes between $0.06\lessapprox a_{\rm amC}\lessapprox0.1$ and $0.11 \lessapprox a_{\rm amC}\lessapprox 0.14$ $\mu$m, respectively. The maximum grain size is attained for the reddest X-stars and is around $\approx 0.16$ $\mu$m.

As discussed in \citet{Nanni16}, the final grain size is mostly dependent on the choice of the seed particle abundance and is expected to roughly follow the relation:
\begin{equation}\label{size_scaled}
a_{\rm i}\propto (\epsilon_{\rm s})^{-1/3},
\end{equation}
where $\epsilon_{\rm s}$ is the adjustable parameter of Eq.~\ref{n_seeds}. 
The grain sizes are therefore expected to be a factor of $\approx 2$ smaller than the ones of R13, if $\log(\epsilon_{\rm s})=-12$ (R12, Z12, J400, J1000), and $\approx 4.6$ smaller, if $\log(\epsilon_{\rm s})=-11$ (H11).
In Fig.~\ref{size_scaled_f} we show the ratios between the grain sizes obtained with R13 and the ones derived from H11 and R12, as a function of \jks\ .
From the figure we conclude that the expected scaling factors between the grain sizes are roughly recovered.

The final grain sizes obtained from our models are clearly different between the various classes of stars and can be also very different from the value of $a_{\rm amC}\approx 0.1$~$\mu$m usually adopted in the literature.

For all the optical data sets considered, the uncertainty associated to the size of carbon grains is usually below few per cents and anyway within $\approx$30$\%$, for the least dust enshrouded stars, while is always around $3-4\%$ for the reddest stars.

\begin{figure}
\includegraphics[trim=0 0 0 0]{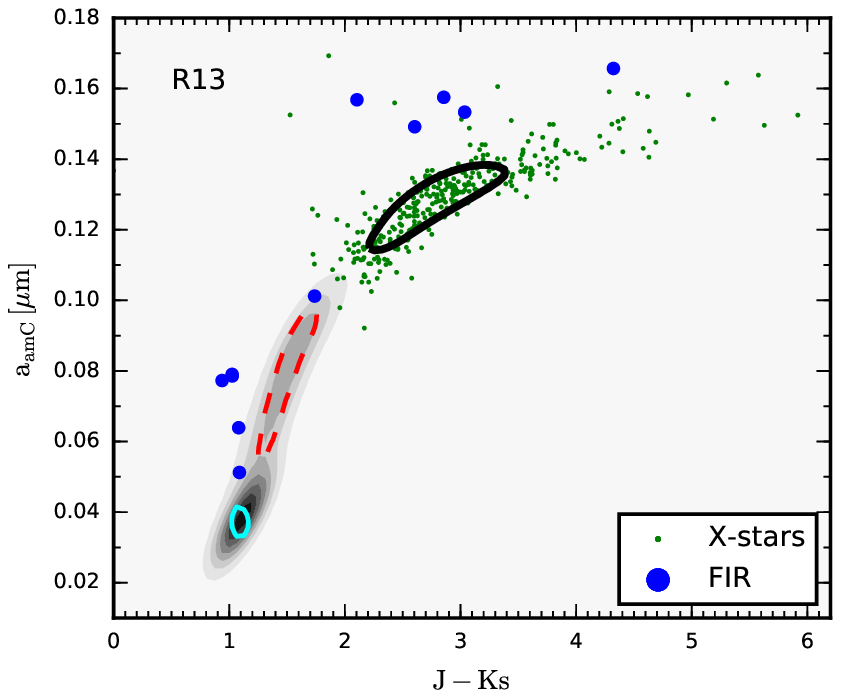}
                \caption{ Size of carbon grains grain as a function of the \jks\ color derived for R13 optical data set. The color code for the different classes of stars is the same as in Fig.~\ref{ml}.}
        \label{rc}
        \end{figure}

\begin{figure}
\includegraphics[trim=0 0 0 0]{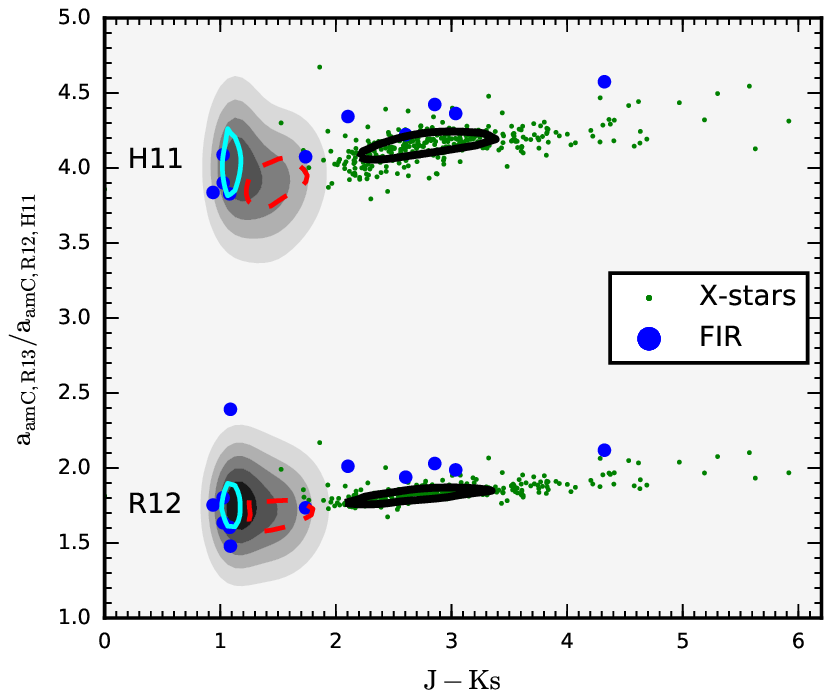}
        \caption{Ratios between the grain sizes obtained with R13 and the ones derived from H11 and R12, as a function of the \jks\ color. The color code for the different classes of stars is the same as in Fig.~\ref{ml}.}
        \label{size_scaled_f}
        \end{figure}

The trend of SiC grain size as a function of \jks\ is qualitatively similar to the one recovered for carbon dust, but the final size of SiC grains are always $\lessapprox0.04$~$\mu$m. 
The reason for this low value of the grain size is that SiC is limited by the silicon abundance.
The uncertainty associated to the SiC grain size is typically between $\approx10\%$ and $\approx40\%$ for the least dust-enshrouded stars and it decreases for the dustiest stars, down to $\approx15\%$.

\subsubsection{Expansion velocities}\label{V_scaling}
In this Section we discuss the expansion velocities obtained by our SED fitting procedure.

In Fig.~\ref{vel_mloss} the expansion velocities against the mass-loss rates are plotted for all the optical data sets. Only the stars for which the outflow is accelerated ($v_{\rm exp} \ge 5$~km~s$^{-1}$) are shown.

For all the data sets, the expansion velocity increases with the mass-loss rate for C-stars and it flattens for X-stars attaining a maximum value around $-5.7\lessapprox\dot{M}\lessapprox-5.6$. The scatter of the velocities of X-stars is always large.
Due to the low amount of dust produced in their CSEs, few aAGBs are accelerated through dust-driven wind, but only for H11, J1000 and J400.
FIR sources show different velocities which are related to the dust content in their CSEs.
The dust-driven wind is sustained for different minimum values of the mass-loss rate for the different optical data sets.

The trends of the expansion velocity and the density distributions of the stars are similar for J1000, Z12 and J400. Among these data sets, the scatter of the velocities of X-stars obtained with J400 is the largest. 
The bulk of C-stars show velocities of $11\lessapprox v_{\rm exp}\lessapprox19$~km~s$^{-1}$ and mass-loss rates $-6.5\lessapprox\log(\dot{M})\lessapprox-6$. For mass-loss rates lower than $\log(\dot{M})\lessapprox-6.5$ ($\lessapprox-6.7$ for J400) the outflow is not efficiently accelerated. Most of X-stars show velocities between $15\lessapprox v_{\rm exp}\lessapprox 21$~km~s$^{-1}$ for J1000 and Z12 and $13\lessapprox v_{\rm exp}\lessapprox 21$~km~s$^{-1}$ for J400 with $-6\lessapprox\log(\dot{M})\lessapprox-5$.
Only few, heavily mass-losing stars, exhibit velocities up to $v_{\rm exp}\approx30$~km~s$^{-1}$.

For R12 data set, the velocities of C-stars are similar to J1000, Z12 and J400, but the bulk of these stars show larger mass-loss rates $-6.2\lessapprox\log(\dot{M})\lessapprox-5.8$. The mass-loss required for accelerating the outflow is also slightly larger with respect to the ones previously discussed.
On the other hand, X-stars reach slightly lower expansion velocities, usually below $\approx20$~km~s$^{-1}$ for $-6\lessapprox\log(\dot{M})\lessapprox-5$. Only few sources reach larger velocities, $v_{\rm exp}\approx27$~km~s$^{-1}$. 

The velocities obtained with R13 are the lowest. For this data sets, the velocities of most of the C-stars are between $10\lessapprox v_{\rm exp}\lessapprox 16$~km~s$^{-1}$ with $-6\lessapprox\log(\dot{M})\lessapprox-5.8$. The minimum mass-loss rate for the wind acceleration is $\log(\dot{M})\approx-6.2$. 
The majority of X-stars shows velocities between $11\lessapprox v_{\rm exp}\lessapprox17$~km~s$^{-1}$. 
Only few X-stars, with the largest mass-loss rates, attain velocities up to $25\lessapprox v_{\rm exp}\lessapprox 27$~km~s$^{-1}$.

The velocities derived with H11 are $12\lessapprox v_{\exp}\lessapprox 24$~km~s$^{-1}$ for most of the C-stars and are concentrated in a range of mass-loss rates broader than the ones of the other data sets, $-6.7\lessapprox\log(\dot{M})\lessapprox-6$. The outflow acceleration starts to be efficient at $\log(\dot{M})\lessapprox-6.9$. The velocities of X-stars are the largest obtained and are concentrated around $19\lessapprox v_{\rm exp}\lessapprox 26$~km~s$^{-1}$. 
The maximum expansion velocity attained (for few sources) is up to $\approx 30-35$~km~s$^{-1}$.

The uncertainty associated to the expansion velocity is usually between $\approx10\%$ and $\approx70\%$ for the least dusty stars, and between $\approx30\%$ and $\approx40\%$ for most of the dustiest sources.

The panels of Fig.~\ref{vel_mloss} are compared with observations, shown in Fig.~\ref{vel_mloss_data}. In this figure, the expansion velocities are plotted against the mass-loss rate derived from observations in our Galaxy \citep[][Table 4, and references therein]{Schoier01,Groenewegen02,RamstedtOlofsson_14, Danilovich_etal15, Groenewegen16}, and in the LMC \citep{Groenewegen16}.
The two samples in \citet{Groenewegen02} and \citet[][Table 4]{Groenewegen16} are infrared luminous, dust rich sources. In particular, the stars in the sample in \citet[][Table 4]{Groenewegen16} have been named as ``Galactic extreme'', as also reported in the figure.
By comparing Fig.~\ref{vel_mloss} and  Fig.~\ref{vel_mloss_data}, we can see that the trend between $v_{\rm exp}$ and $\dot{M}$ derived for carbon stars in the SMC is quite similar to the one observed for Galactic sources. 

The observed velocities of Galactic stars exhibit quite a large scatter, especially at the largest mass-loss rates. This effect is also predicted by our analysis for the SMC carbon stars.
In the range of mass-loss rates $-7\lessapprox\log\dot{M}\lessapprox-6.5$ the stars observed in our Galaxy already show expansion velocities of $6\lessapprox v_{\rm exp}\lessapprox10$~km~s$^{-1}$. Similar velocities in the same range of mass-loss rates are only obtained for H11 in the SMC. Most of the Galactic stars, exhibit velocities around $10\lessapprox v_{\rm exp}\lessapprox20$~km~s$^{-1}$ for mass-loss rates of $-5.5\lessapprox\log\dot{M}\lessapprox-4.5$. These values are also attained by SMC stars in a similar range of mass-loss, with the exclusion of R13 data set, for which lower velocities are predicted. In the same range of mass-loss rates, some Galactic stars reach velocities between $20\lessapprox v_{\rm exp}\lessapprox25$~km~s$^{-1}$, which are only achieved by SMC sources fitted with H11.
The largest velocities observed for Galactic stars, $v_{\rm exp}\approx40-45$~km~s$^{-1}$, are never attained for the SMC stars.

Excluding the Galactic stars with $v_{\rm exp}\gtrapprox 25$~km~s$^{-1}$ and the ones with $-7\lessapprox\log\dot{M}\lessapprox-6.5$, the observed velocities of C-rich stars in our Galaxy are in general comparable to the ones derived for the SMC. This finding indicates that there is not a large difference in the predicted velocities of carbon stars for different metallicity. Our result is in line with the analysis performed by means of hydrodynamical simulations  by \citet{Mattsson10} and \citet{Eriksson14}. These authors concluded that the carbon-excess, rather than metallicity, is one of the key parameters which determines the properties of the dust-driven wind of carbon stars. In particular, the expansion velocities predicted by theoretical models is expected to be dependent on the choice of the carbon excess, with larger expansion velocities for larger values of carbon-excess.

\begin{figure*}
\includegraphics[trim= 1cm 0 0 0]{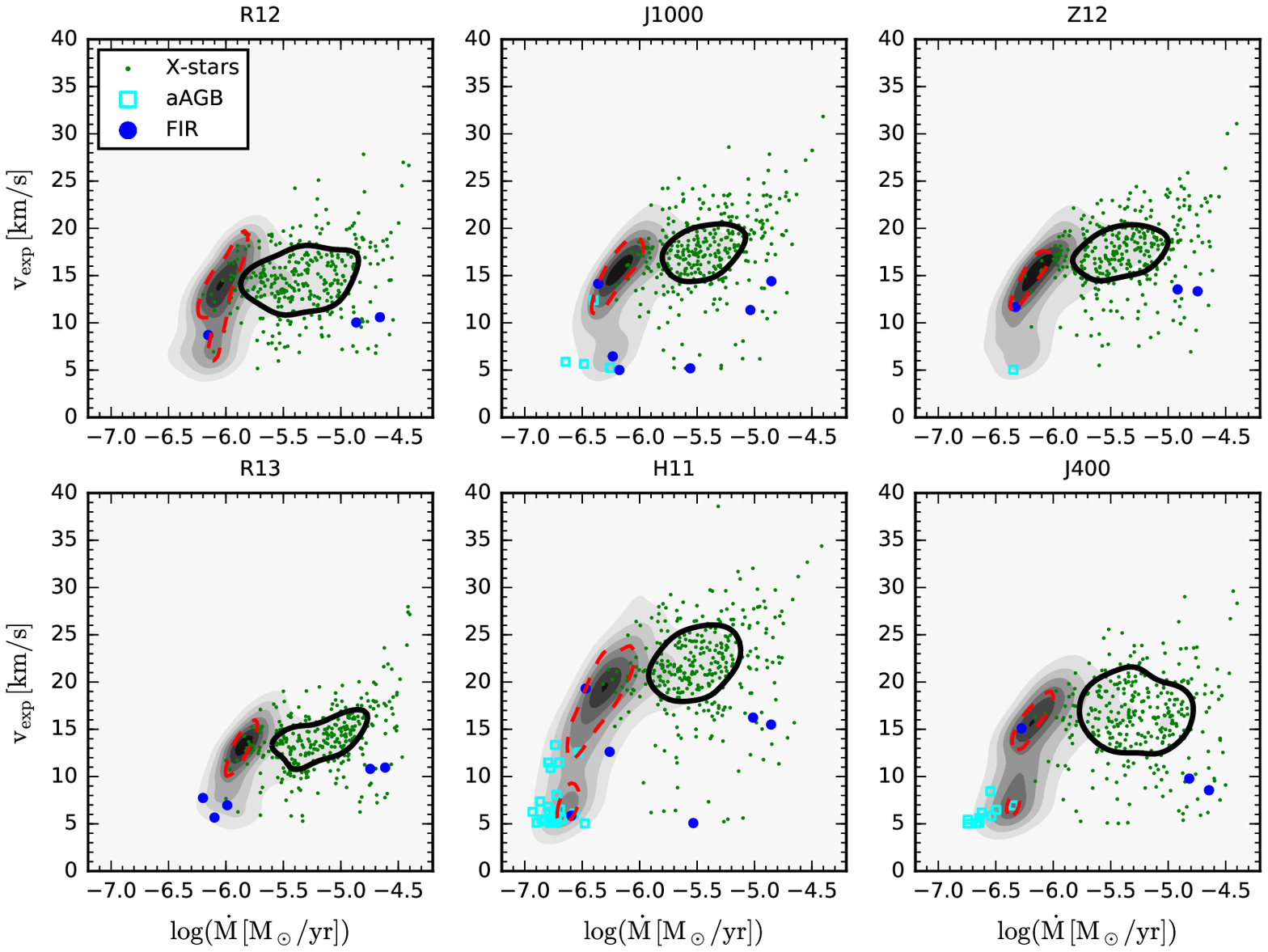}
        \caption{Expansion velocity as a function of the mass-loss rate for all the optical data sets of carbon dust. The color code for the different classes of stars is the same as Fig.~\ref{ml}.}
        \label{vel_mloss}
        \end{figure*}

        \begin{figure}
\includegraphics[trim=0 0 0 0]{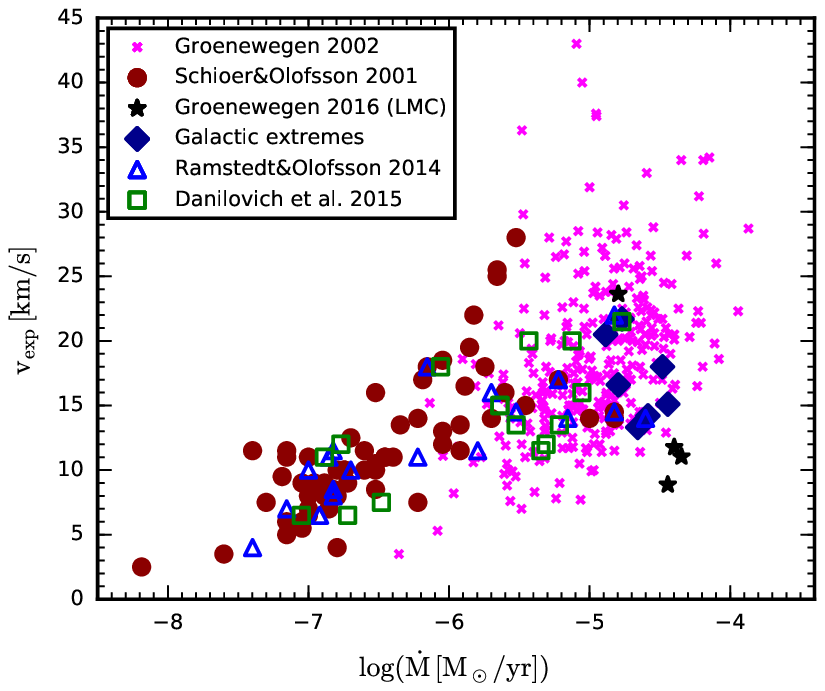}
        \caption{Expansion velocity as a function of the mass-loss rate  of carbon stars in our Galaxy derived by different authors \citep[][Table 4]{Schoier01,Groenewegen02,RamstedtOlofsson_14, Danilovich_etal15, Groenewegen16}, and in the LMC \citep{Groenewegen16}, represented with different colors and symbols listed in the legend. Stars named as ``Galactic extreme'' are taken from Table 4 of \citet{Groenewegen16}.}
        \label{vel_mloss_data}
        \end{figure}

In Fig.~\ref{vel_contour} we plot the expansion velocities ($v_{\rm exp} \ge 5$~km~s$^{-1}$) as a function of the luminosity.
For comparison, we show the expansion velocities obtained from the scaling relation in Eq.~\ref{vexp_scaled} (black squares), adopted in \citet{Boyer12} and by \citet{Srinivasan16} with the assumption $v_{\rm exp}=10$~km~s$^{-1}$ for $L=30000$~$L_\odot$ and $\Psi_{\rm dust}=200$. 
The expansion velocity correlates with the luminosity, as predicted by the theory of dust-driven wind \citep{Elitzur01,Ivezic10}. 
For all the data sets, the outflow acceleration occurs at about the same value of the luminosity, $L\approx 2000$ L$_{\odot}$. The linear trend between the velocity and luminosity is particularly evident for the dustiest stars, which reach their maximum expansion velocity for the largest luminosity, $L\approx25000$~$L_{\odot}$.

In Fig.~\ref{vel_data} we plot the observed velocities as a function of the luminosity for the same Galactic and LMC samples of Fig. \ref{vel_mloss_data}.
By comparing Figs.~\ref{vel_contour} and~\ref{vel_data} we can immediately see that the trend between the velocity and luminosity for the SMC stars, is similar to the one observed for Galactic stars. 
Some of the observed Galactic sources have $2\lessapprox v_{\rm exp}\lessapprox10$~km~s$^{-1}$ just above 2000~$L_{\odot}$ \citep{Schoier01}. This result is usually in line with the one obtained for SMC stars.
Around $L=6000$~$L_\odot$, the observed velocities of Galactic stars are $6\lessapprox v_{\rm exp}\lessapprox18$~km~s$^{-1}$. These values are similar to the ones of SMC stars for all the data sets.
The bulk of the Galactic stars with the largest velocities is around $8000 L_{\odot} \lessapprox L \lessapprox9000$~$L_{\odot}$. At these luminosities the velocities attained by SMC stars are between $20 {\rm km~s}^{-1},\lessapprox v_{\rm exp}\lessapprox25$~km~s$^{-1}$, similarly to most of the Galactic stars. Lower velocities are expected by adopting R13, whilst by employing H11 the velocities are up to $v_{\rm exp}\approx28$~km~s$^{-1}$.
Velocities up to $v_{\rm exp}\approx40-45$~km~s$^{-1}$ are reached for some of the Galactic stars at $8000 L_{\odot} \lessapprox L\lessapprox10000$~$L_{\odot}$, whereas the maximum value reached by our models is $v_{\rm exp}\approx39$~km~s$^{-1}$ for higher luminosities $L\approx25000$~$L_{\odot}$ and only with H11. 

We again find that the dynamical properties of SMC carbon stars are in general comparable to the ones of Galactic sources and they are not significantly related to the initial metallicity.

In Fig.~\ref{velvsL_psi} we plot the average velocities of the stars as a function of the luminosity, color coded with $\log\Psi_{\rm dust}$ for R12 data set. 
The results for a selected value of $\Psi_{\rm dust}\approx 800$ are also overplotted.
The general trends recovered are analog for different data sets.
For a given luminosity, larger velocities are found at increasing dust content, corresponding to lower values of $\Psi_{\rm dust}$, as also assumed in Eq.~\ref{vexp_scaled} and predicted by the wind theory.
However, for the plotted value of $\Psi_{\rm dust}=800$ and for a given luminosity, the scatter of the velocities is up to $\approx7-8$~km~s$^{-1}$. 
The predicted spread is expected since other quantities, such as the T$_{\rm inn}$, T$_{\rm eff}$ and $\dot{M}$, change among the plotted sources, affecting the velocities \citep{Elitzur01, Ivezic10}.
A clear trend between $v_{\rm exp}$ and $\Psi_{\rm dust}$ is highlighted if the variation in $\Psi_{\rm dust}$ spans a large range of values, which is more than one order of magnitude in Fig.~\ref{velvsL_psi}.

In the upper panel of Fig.~\ref{vel_ratio} we plot the ratios between the velocities obtained with R13 and the value adopted in \citet{Groenewegen09} as a function of \jks. We select for the comparison our R13 data set, since \citet{Groenewegen09} adopted \citet{Rouleau91} in their calculations, assuming that all grains have size of $a\approx0.1$~$\mu$m. The optical properties in \citet{Groenewegen09} are calculated for a continuous distribution of hollow spheres and not for simple spherical grains (Mie theory).
The velocities we predict for the least dust-rich stars are typically less than one half of the assumed value of 10~km~s$^{-1}$. For these stars the dust-driven wind is expected to be inefficient with R13.
On the other hand, the typical velocities of the dustiest stars are between $\approx1.1-1.7$ times larger than the assumed ones.

In the lower panel of Fig.~\ref{vel_ratio}, the expansion velocities obtained for Z12 are compared with the one assumed by \citet{Srinivasan16} who adopted the ACAR data set by \citet{Zubko96} with typical spherical grain size of $a\approx0.1$~$\mu$m.
Similarly to the upper plot, the bulk of the dust-poor stars are not efficiently accelerated and show velocities lower than the assumed ones. On the other hand, the predicted velocities of the most dust-rich sources are between $\approx1.9-2.7$ times larger than the assumed ones.

\begin{figure*}
\includegraphics[trim= 1cm 0 0 0]{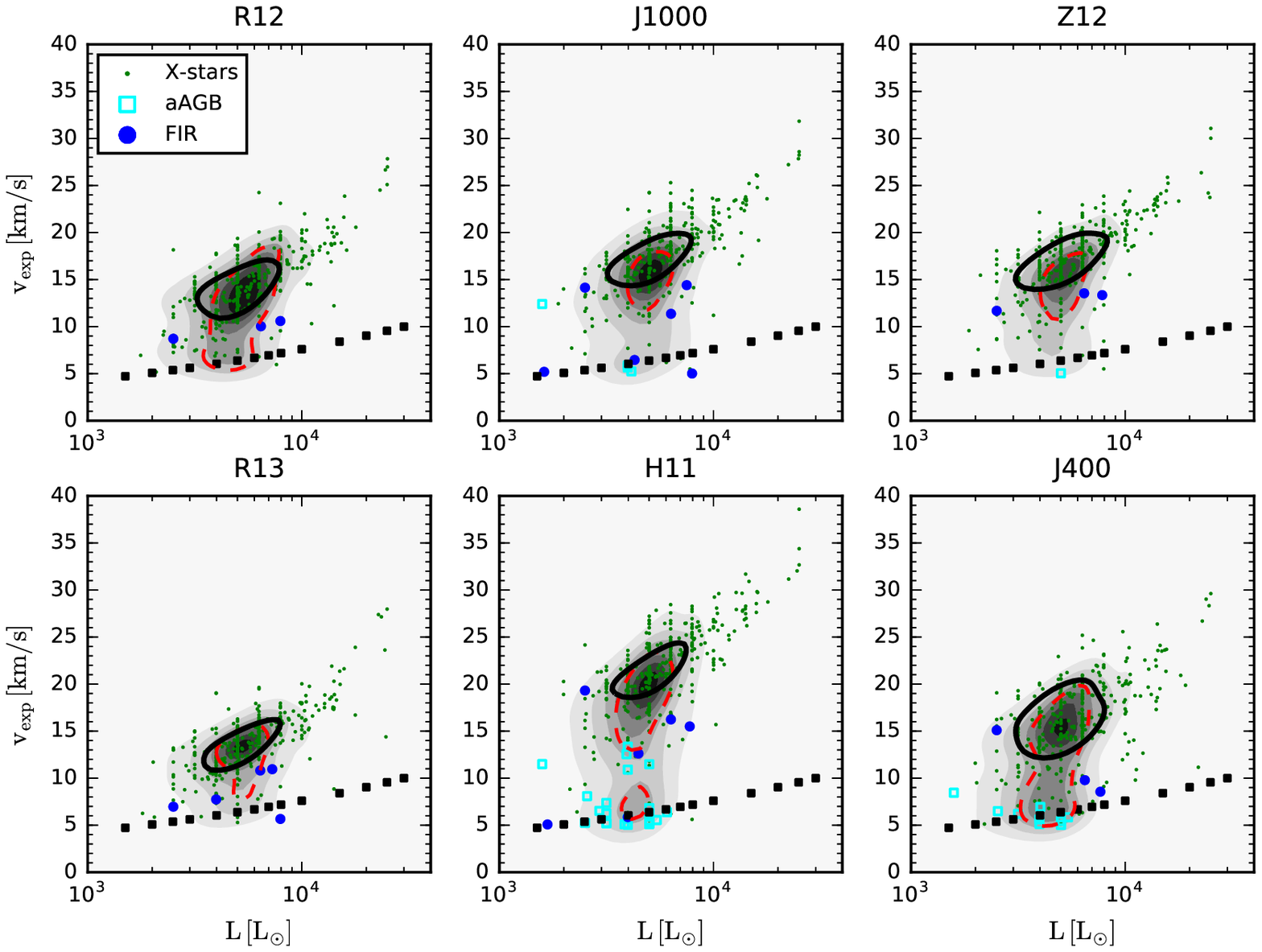}
\caption{Expansion velocity as a function of the stellar luminosity for all the optical data sets of carbon dust. For comparison, the expansion velocities derived from the scaling relation in Eq.~\ref{vexp_scaled}, as in \citet{Boyer12} and \citet{Srinivasan16} are also plotted with full black squares (see text for more details). The color code for the different classes of stars is the same as Fig.~\ref{ml}.}
        \label{vel_contour}
        \end{figure*}

\begin{figure}
\includegraphics[trim= 0 0 0 0]{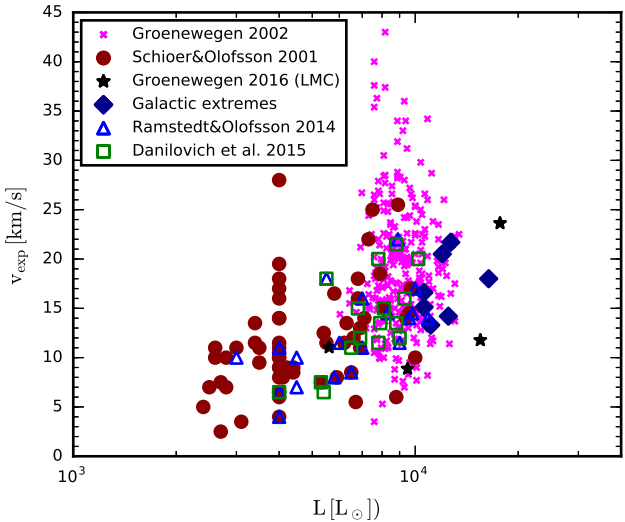}
        \caption{Expansion velocity as a function of the luminosity for the same observations of Galactic and LMC C-rich sources shown in Fig.~\ref{vel_mloss_data}.}
        \label{vel_data}
        \end{figure}

    \begin{figure}
\includegraphics[trim=0 -1 0 0]{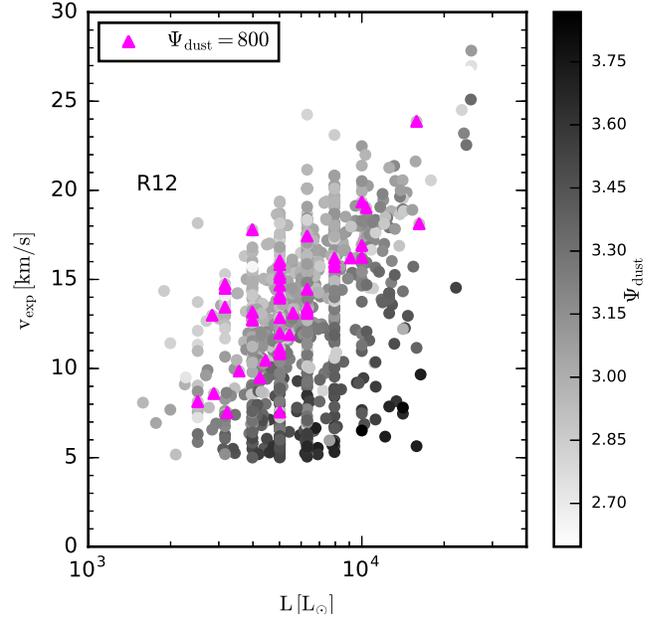}
        \caption{Expansion velocity as a function of luminosity color coded for the value of $\Psi_{\rm dust}$, derived for R12 data set. The stars with $\Psi_{\rm dust}\approx 800$ are overplotted (magenta triangles).}
        \label{velvsL_psi}
        \end{figure}
        
            \begin{figure}
\includegraphics[trim=0 0 0 1cm]{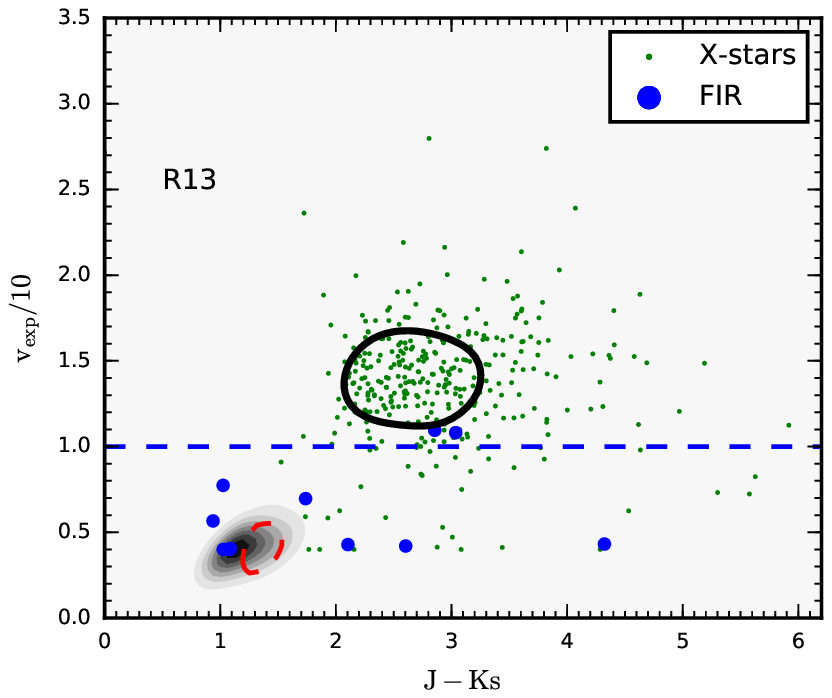}
\includegraphics[trim=0 0 0 0]{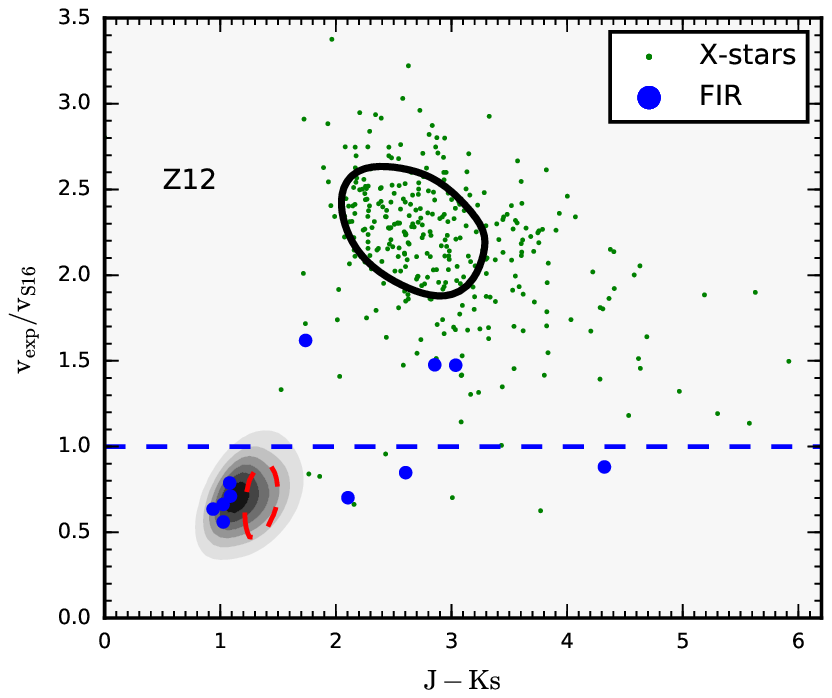}
        \caption{Upper panel: ratio between the velocities obtained from the SED fitting procedure by employing R13 data set, and $v_{\rm exp}=10$~km~s$^{-1}$, as assumed by \citet{Groenewegen09},  as a function of the \jks\ color. Lower panel: the same as in the upper panel, but selecting the results for Z12 data set and computing $v_{\rm exp}$ for all the stars by employing the scaling relation in Eq.~\ref{vexp_scaled}, as assumed by \citet{Srinivasan16}. The different classes of stars are plotted with the same color code as in Fig.~\ref{ml}.}
        \label{vel_ratio}
        \end{figure}

\subsubsection{Carbon excess}
On the base of the analysis presented we do not find a clear trend between the value of the carbon excess and \jks. This can be explained by the small dependence usually found between $\tau_1$ and the carbon excess compared to other stellar parameters, as the mass-loss rate, as shown in the Section~\ref{trends_tau}. 
As a consequence, the carbon excess is not well constrained and the values considered in our grids are almost equally probable for the majority of the sources for all the data sets considered.
Different combinations of the mass-loss rates and carbon excess produce a valid fit of the selected source. In particular, for larger values of the mass-loss rate, the corresponding carbon excess increases, whilst for lower values of the mass-loss rate the carbon excess increases.
Such a result suggest that there is a degeneracy between the mass-loss rate and the carbon excess.

\subsubsection{Current stellar mass}
As well known, the current stellar mass has a minor effect on the emerging stellar spectra.
For this reason, we are unable to constrain the current stellar mass by our analysis similarly to what we find for the carbon excess.
A better constrain on the current stellar mass might be provided from independent observational determination of the expansion velocities and mass-loss rates of carbon stars in the SMC. 

\subsection{Dust production rates}
We here discuss the DPRs derived for the different optical data sets selected for carbon dust.
Our DPRs are shown for the various dust species formed (amC, SiC and iron) and for the different classes of stars, photometrically classified. 
For aAGBs, we report only half of the DPRs derived from the fitted sources, since about half of the aAGBs are carbon-rich \citep{Boyer15}.
Our DPRs are compared with the ones found in the literature.
\subsubsection{DPRs for the different dust species}
The DPRs for the different dust species and optical data sets for carbon dust are listed in Table~\ref{DPR_i}. 
For all the optical data sets, the total amount of dust produced is dominated by carbon dust for all the type of stars. 
The amount of iron dust is low and very uncertain. The smallest uncertainty in the iron dust production is found for the most dust poor stars and it is between $\approx50-72\%$ depending on the optical data set.
The total amount of SiC condensed is as well rather uncertain for the most dust poor stars. The lowest uncertainties are between $\approx33$ and $\approx56\%$, depending on the data set, for X-stars and FIR.
The large uncertainties affecting the amount of SiC and iron dust is not surprising since these dust species are affecting the SED of carbon-rich stars less than carbon dust. 
Moreover, as discussed in Section \ref{sec:dtg}, a larger uncertainty in the gas-to-dust ratio is expected for the least dusty stars, for which we expect a larger degeneracy in the parameters producing a spectrum which is almost dust-free.

Most of the SiC is produced in CSEs of X-stars. The total amount of SiC can vary up to a factor $\approx 2.7$ for the different optical data sets (see J1000 and J400). The SiC mass fraction over the total is between $\approx 4$ and $\approx8\%$ (see R13 and H11).
On the other hand, the total iron dust produced is always negligible.

\begin{table*}
\begin{center}
\caption{DPRs in $M_{\odot}$yr$^{-1}$ divided by dust type and stellar classes. The DPRs listed are computed for the different combinations of optical data sets selected.}
\label{DPR_i}
\begin{tabular}{c l l l}
\hline
R13& Carbon    &        SiC     &   Iron  \\
\hline
C-stars   & $(5.37\pm0.78)\times10^{-7}$  &   $(1.13\pm0.86) \times10^{-9}$  & $(6.4\pm4.2)\times10^{-11}$  \\  
X-stars  & $(2.72\pm 0.82)\times10^{-6}$  &    $(1.47\pm0.70)\times10^{-7}$  &   $(3.1\pm3.3)\times10^{-10}$\\  
aAGB  & $(6.8\pm3.4)\times10^{-9}$  &     $(3.1\pm2.6)\times10^{-11}$  &     $(4.7\pm3.3)\times10^{-12}$  \\  
FIR &  $(6.8\pm2.9)\times10^{-8}$  &     $(4.9\pm2.7)\times10^{-9}$ & $(2.3\pm2.4)\times10^{-11}$  \\  
Total &  $(3.34\pm0.93)\times10^{-6}$  & $(1.53\pm0.73)\times10^{-7}$  &  $(4.0\pm 4.0)\times10^{-10}$\\
Mass fraction   &   $\approx96\%$  &       $\approx4\%$  &     - \\
\hline
R12& & & \\
\hline
C-stars &  $(6.7\pm1.1)\times10^{-7}$  &     $(2.3\pm1.5)\times10^{-9}$  &  $(1.62\pm0.92)\times10^{-11}$  \\  
X-stars  & $(2.98\pm0.97)\times10^{-6}$  &    $ (2.03\pm0.86)\times10^{-8}$  &   $(2.2\pm2.5)\times10^{-9}$  \\  
aAGB  & $(7.4\pm3.7)\times10^{-9}$  &     $(4.5\pm3.3)\times10^{-11}$  &     $(9.3\pm 5.8)\times10^{-12}$  \\  
FIR  & $(6.8\pm2.9)\times10^{-8}$  &     $(6.3\pm2.5)\times10^{-9}$  &     $(1.9\pm  2.1)\times10^{-10}$  \\ 
Total &  $(3.7\pm 1.1)\times10^{-6}$  &     $(2.12\pm0.90)\times10^{-7}$ &  $(2.5\pm   2.8)\times10^{-9}$\\
Mass fraction   & $\approx95\%$  &       $\approx5\%$  &     -\\
\hline
J1000& & & \\
\hline
C-stars &  $(2.57\pm0.50)\times10^{-7}$  &     $(9.0\pm 6.3)\times10^{-10}$  &  $(5.8\pm3.4)\times10^{-11}$  \\  
X-stars  & $(1.30\pm0.37)\times10^{-6}$   &     $(9.6\pm4.3)\times10^{-8}$  &     $(2.1\pm1.7)\times10^{-10}$  \\
aAGB  & $(2.4\pm1.3)\times10^{-9}$  &     $(1.7\pm1.3)\times10^{-11}$  &     $(4.2\pm2.7)\times10^{-12}$  \\  
FIR  & $(2.9\pm1.2)\times10^{-8}$  &     $(2.7\pm1.5)\times10^{-9}$  &     $(1.22\pm0.96)\times10^{-11}$  \\  
Total  & $(1.58\pm0.44)\times10^{-6}$  &     $(1.00\pm 0.45) \times10^{-7}$& $(2.8\pm   2.2)\times10^{-10}$\\
Mass fraction   &   $\approx94\%$  &       $\approx6\%$  &     -\\
\hline
J400& & & \\
\hline
C-stars &  $(4.9\pm1.3)\times10^{-7}$  &     $(1.5\pm1.3)\times10^{-9}$  &  $(6.5\pm4.6)\times10^{-11}$  \\  
X-stars &  $(3.2\pm1.1)\times10^{-6}$  &     $(2.6\pm1.0)\times10^{-7}$  &     $(4.3\pm  4.4)\times10^{-9}$  \\  
aAGB  & $(4.5\pm2.8)\times10^{-9}$  &     $(1.8\pm1.5)\times10^{-11}$  &     $(4.8\pm   3.2)\times10^{-12}$  \\  
FIR &  $(7.0\pm3.1)\times10^{-8}$  & $(7.2\pm2.9)\times10^{-9}$  &  $(3.2\pm 3.3)\times10^{-10}$  \\ Total  & $(3.8\pm1.2)\times10^{-6}$  &     $(2.7\pm1.1)\times10^{-7}$  &     $(4.7\pm4.8)\times10^{-9}$\\
Mass fraction   &   $\approx93\%$  &      $\approx7\%$  &      -\\
\hline
Z12& & & \\
\hline
C-stars  & $(3.30\pm0.62\times10^{-7}$  &   $(1.0\pm0.76)\times10^{-9}$  &  $(6.2\pm3.9)\times10^{-11}$\\  
X-stars &  $(1.83\pm0.54)\times10^{-6}$  &  $(1.42\pm0.62)\times10^{-7}$  & $(3.9\pm3.7)\times10^{-10}$  \\  
aAGB &  $(3.3\pm1.7)\times10^{-9}$  &  $(2.1\pm1.6)\times10^{-11}$  &  $(5.2\pm3.1)\times10^{-12}$  \\  
FIR &  $(4.4\pm1.7)\times10^{-8}$  &     $(4.4\pm2.0)\times10^{-9}$  &     $(2.9\pm2.6)\times10^{-11}$  \\  
Total &  $(2.20\pm0.63)\times10^{-6}$  &     $(1.47\pm0.65)\times10^{-7}$  & $(4.8\pm4.4)\times10^{-10}$\\
Mass fraction  &    $\approx94\%$  &      $\approx6\%$  &     -\\
\hline
H11& & & \\
\hline
C-stars &  $(3.49\pm0.81)\times10^{-7}$  &  $(2.7\pm1.5)\times10^{-9}$ & $(1.23\pm   0.72)\times10^{-10}$  \\  
X-stars &  $(1.72\pm0.51)\times10^{-6}$  &  $(1.65\pm0.58)\times10^{-7}$  & $(1.2\pm   1.1)\times10^{-9}$  \\  
aAGB &  $(2.8\pm1.5)\times10^{-9}$  &     $(2.8\pm1.8)\times10^{-11}$  &    $(1.0\pm   0.51)\times10^{-11}$  \\  
FIR &  $(3.7\pm1.5)\times10^{-8}$  &     $(4.2\pm1.4)\times10^{-9}$  &     $(8.4\pm 7.5)\times10^{-11}$  \\  
Total &  $(2.11\pm0.61)\times10^{-6}$  &  $(1.72\pm 0.61)\times10^{-7}$ &  $(1.4\pm   1.3)\times10^{-9}$\\
Mass fraction  &    $\approx92\%$  &       $\approx 8\%$  &    -\\
\hline
\end{tabular}
\end{center}
\end{table*}

\subsubsection{Contribution of the different classes of stars to the total DPR}
In Table~\ref{DPR} the values of the DPRs are shown for the different classes of stars.
As far as aAGBs are concerned, we list as final DPR half of the total amount obtained from our analysis, since, as discussed by \citet{Boyer15}, only about half of the aAGBs are expected to be C-rich.

The largest contribution to the total DPR is always provided by X-stars ($\approx 82-87\%$) and the remaining DPR is mainly due to C-stars. In fact, aAGBs always yield a very small amount of dust with respect to the total ($\approx 0.1\%$).
The DPR of the FIR sources is usually larger than the one of all the aAGB stars, even though the value is rather uncertain.
The DPR of FIR is only $\approx2\%$ of the DPRs of all the other sources.
However, the amount of dust produced by the FIR is surprisingly large if compared with the one of aAGBs, since FIR are only 11, while aAGBs are $\approx550$.
The average DPR of FIR is about $\approx3-7\times10^{-9}$~$M_{\odot}$~yr$^{-1}$, depending on the optical data set.  This value is close to the one of X-stars and is typically $\approx20$ times larger than the one of C-star. On the other hand, the average DPR of aAGBs is typically $\approx 10^{-11}$~$M_{\odot}$~yr$^{-1}$ or less.

\subsubsection{Total DPRs for the different optical data sets and compared with the literature}
\begin{table*}
\caption{Total and average DPRs in $M_{\odot}$yr$^{-1}$ computed for the different combinations of optical data sets selected and listed for the different classes of stars. Other DPRs found in the literature are also shown.}
\label{DPR}
\begin{center}
\begin{tabular}{c l l l l l}
\hline
This work, for dataset:  & C-stars  &   X-stars  &  aAGB  &  FIR & Total (no FIR)   \\ 
  Number of stars & 1709 & 339 & 1092 & 11 &-\\
R13&    $(5.38\pm0.79)\times10^{-7}$ &   $(2.87\pm0.89)\times10^{-6}$&   $(6.9\pm   3.4)\times10^{-9}$ &   $(7.3\pm3.2)\times10^{-8}$ &   $(3.42\pm0.98)\times10^{-6}$\\
R12&    $(6.7\pm1.1)\times10^{-7}$ &   $(3.2\pm1.1)\times10^{-6}$ &  $(7.4\pm   3.7)\times10^{-9}$ &   $(7.4\pm 3.2)\times10^{-8}$&   $(3.9\pm1.2)\times10^{-6}$ \\
J1000&    $(2.57\pm0.50)\times10^{-7}$ &   $(1.39\pm0.42)\times10^{-6}$ &   $(2.4\pm   1.3)\times10^{-9}$ &   $(3.2\pm1.3)\times10^{-8}$&   $(1.65\pm0.47)\times10^{-6}$\\
J400&    $(5.0\pm1.3)\times10^{-7}$&   $(3.5\pm1.2)\times10^{-6}$&   $(4.5\pm   2.8)\times10^{-9}$&   $(7.8\pm3.4)\times10^{-8}$&   $(4.0\pm1.3)\times10^{-6}$\\
Z12&    $(3.31\pm0.63)\times10^{-7}$  &   $(1.97\pm0.61)\times10^{-6}$  &   $(3.3\pm   1.7)\times10^{-9}$ &   $(4.9\pm1.9)\times10^{-8}$ &   $(2.30\pm0.67)\times10^{-6}$ \\
H11&    $(3.52\pm0.82)\times10^{-7}$&   $(1.89\pm0.57)\times10^{-6}$&   $(2.8\pm   1.5)\times10^{-9}$&   $(4.2\pm1.6)\times10^{-8}$ &   $(2.24\pm0.66)\times10^{-6}$  \\
\hline
\citet{Srinivasan16}  & C-stars  &   X-stars  &  aAGB  &  FIR & Total (no FIR)   \\ 
 Number of stars & 1652 & 337 & - & - & -\\
  & $\approx$1.2$\times 10^{-7}$ &  $\approx$6.8$\times 10^{-7}$ & - & - & $\approx$8.0$\times 10^{-7}$ \\
\hline
\citet{Boyer12}  & C-stars  &   X-stars  &  aAGB  &  FIR & Total (no FIR)   \\ 
Number of stars & 1559 & 313  & - & - &- \\
  & $\approx$1.2$\times 10^{-7}$ &  $\approx$6.3$\times 10^{-7}$ & - & - & $\approx$7.5$\times 10^{-7}$\\
\hline
\citet{Matsuura13}& - & - & - & - & $\approx$4$\times 10^{-6}$\\
\hline
\end{tabular}
\end{center}
\end{table*}

The total DPRs obtained for different optical data sets are shown in Table~\ref{DPR} together with other results found in the literature.
Depending on the optical data set, we find that the first $\sim$250 most dust producing stars provide the $\approx$80$\%$ of the total DPR. 
The two most extreme values of the DPRs are obtained for J1000 ($\approx 1.65\times10^{-6}$~M$_\odot$~yr~$^{-1}$) and J400 ($\approx 4.0\times10^{-6}$~M$_\odot$~yr~$^{-1}$), respectively. The variation between these two values is of a factor $\approx2.4$. 

For all the optical data sets, the typical uncertainty of the total dust production is around $\approx30\%$.
Since the uncertainties were estimated by analyzing the variation of the photometry due to a random variation within the photometric error, we conclude that the precision of the photometry allows for a determination of the total DPR which is typically no better than $\approx30\%$.
Within the estimated uncertainties, the DPRs for the different data sets are compatible except for the one computed with J1000. 

We compare the results of our investigations with others in the literature. 
In the works by \citet{Boyer12} and \citet{Srinivasan16} the optical data set adopted for the SED fitting procedure is the ACAR sample by \citet{Zubko96}. Differently from our approach, a grain size distribution is assumed, with a typical grain size of $a\approx0.1$~$\mu$m. The grains are also assumed to be spherical.
The most natural comparison between \citet{Boyer12}, \citet{Srinivasan16} and our work is with Z12 (see Table~\ref{opt}).
Our DPRs can only be compared with the DPRs of \citet{Boyer12} and \citet{Srinivasan16} computed for C- and X-stars. In fact, in \citet{Boyer12} and \citet{Srinivasan16} a fraction of aAGBs and FIR sources are treated as O-rich and the DPR is computed for the complete sample including C- and O-rich stars. For this reason, the value of the DPRs of aAGBs and FIR for \citet{Boyer12} and \citet{Srinivasan16} are not listed in Table~\ref{DPR}.

From Table~\ref{DPR} we can appreciate that the DPRs computed by \citet{Boyer12}, \citet{Srinivasan16} are $\approx3$ times lower than Z12 for both C- and X-stars.
The average DPRs computed for Z12 are also in line this result. By employing Z12, our average DPR for C-stars is $\approx 1.9\times 10^{-10}$~M$_\odot$~yr~$^{-1}$, to be compared with \citet{Boyer12}, $\approx 7.8\times 10^{-11}$~M$_\odot$~yr~$^{-1}$, and with \citet{Srinivasan16}, $\approx 7.1\times 10^{-11}$~M$_\odot$~yr~$^{-1}$. 
On the other hand, the average DPR of X-stars with Z12 is $\approx5.8\times 10^{-9}$~M$_\odot$~yr~$^{-1}$, which is again almost three times larger than the one derived by \citet{Boyer12} and \citet{Srinivasan16}, $\approx2.0\times 10^{-9}$~M$_\odot$~yr~$^{-1}$.
Furthermore, the DPRs of \citet{Boyer12} and \citet{Srinivasan16} are about $\approx5$ times lower than the largest DPR obtained employing our grid of models (J400). Our lowest DPR, obtained with J1000, is still a factor of $\approx2$ larger than the ones estimated by \citet{Boyer12} and \citet{Srinivasan16}.

The differences between our DPRs and the ones derived by \citet{Boyer12} and \citet{Srinivasan16} for comparable dust opacities are dependent on several factors. First of all, as shown in Fig.~\ref{vel_contour} and in the lower panel of Fig~\ref{vel_ratio} there are considerable differences in the expansion velocities, up to a factor three for the most dust-enshrouded, X-stars. 
The final velocity affects the estimate of the dust mass-loss rate as can be seen from Eq.~\ref{dpr_appr}. 
The possible differences arising as consequence of different assumptions on the expansion velocity, was already pointed out by \citet{Srinivasan16} when comparing their results with the ones by \citet{Matsuura13}.
In addition to that, we do not expect to obtain the same condensation radius, R$_{\rm c}$, from our model and from the fitting procedure by \citet{Srinivasan16}. 
Moreover, as shown in Section \ref{sec:grains} carbon stars fitted with our models can yield grain size quite different from the standard assumption of $0.1$ $\mu$m, which also affects the absorption, scattering and extinction properties of dust grains.

On the other hand, \citet{Matsuura13} estimated the total DPR for carbon-rich stars basing their analysis on the work by \citet{Groenewegen09}. 
In \citet{Groenewegen09}, the optical data set selected for carbon dust is the one by \citet{Rouleau91}. A single value of the grain size of $0.1$~$\mu$m is also adopted. In \citet{Groenewegen09} the optical properties of dust grains are not calculated for spherical grains, but for a continuous distribution of hollow spheres. This choice accounts more consistently for the possible porosity of dust grains, which is not taken into account assuming spherical grains.
The most natural choice is then to compare the total DPR derived by \citet{Matsuura13} with R13, even though we employ spherical grains, for which the optical properties are calculated with the Mie theory.
The total DPR estimated by \citet{Matsuura13} is in fair agreement with {\rm both} R13 and R12 within the uncertainty.
The DPR of \citet{Matsuura13} is however $\approx2.4$ times larger than the lowest value predicted by our analysis (J1000). Furthermore, we never obtain a DPR larger than the one of \citet{Matsuura13} for any of the data sets considered.
        
\section{Summary and conclusions}
In this work we provide physically grounded dusty models and spectra suitable to fit the photometry of the C-rich TP-AGB stars of the SMC.
We perfom such an investigation for some selected choices of optical data sets for carbon dust and seed particle abundances which have been shown to well reproduce simultaneously most of the CCDs in the SMC \citep{Nanni16}.
From the SED fitting procedure we are able to consistently derive some important dust properties and stellar quantities, such as  bolometric luminosities, mass-loss rates, gas-to-dust ratios,  outflow expansion velocities and the dust grain sizes.
These results may be helpfully employed to test the predictions of stellar evolution models \citep{marigoetal13, Marigo_etal16} and synthetic stellar populations including AGB stars \citep{Marigo_etal17}.

We here summarize the main results:
\begin{itemize}
\item \textit{Luminosities}. We find that the luminosities are well constrained by the SED fitting procedure. Our luminosity function is in excellent agreement with the one derived by \citet{Srinivasan16}. The position of the peak is around M$_{\rm bol}\approx-4.5$, in good agreement with the results obtained from the catalog by \citet{Rebeirot_etal93}.
\item \textit{Mass-loss rates}. C- and X-stars are separated in term of mass-loss rates around  $\log\dot{M}\approx-6$, which correspond to $\jks\approx 2$. 
Anomalous AGBs are always characterized by $\log\dot{M}\lessapprox -6$  for all the data sets considered.
The mass-loss rates can vary up to a factor of $\approx2$ for different optical data sets for the most dust-enshrouded X-stars and up to a factor $\approx7-8$ for C-stars, in the most extreme cases.
Our distribution of mass-loss rates for X-stars is similar in terms of shape to the one derived from the observations of dust-enshrouded Galactic stars by \citet{Groenewegen02}, especially for R13 and J400 data sets. However, the position of the peak is shifted to lower mass-loss rates for our distributions.
On the other hand, the distribution of C-stars is compared to the one derived from the Galactic observations by \citet{Schoier01}, for a sample of optically bright carbon stars. Our distributions are more peaked than the one of optically bright carbon stars with the peak shifted to larger values of the mass-loss rate.
%}.
\item \textit{Gas-to-dust ratios}. The majority of X-stars have gas-to-dust ratios lower than C-stars. The value of $\Psi_{\rm dust}$ is never as low as the value usually adopted in the literature ($\Psi_{\rm dust}= 200$). The range of possible values of $\Psi_{\rm dust}$ is quite large, with $500\lessapprox\Psi_{\rm dust}\lessapprox800-2000$, for X-stars, and $\Psi_{\rm dust}\gtrapprox800-2000$, for C-stars. The large values of $\Psi_{\rm dust}$ for aAGBs is consistent with a low amount of dust in the CSEs of these stars, in agreement with \citet{Boyer11}. C- and X-stars are separated around $\Psi_{\rm dust}\approx 800-2000$, depending on the optical data set for carbon dust. 
The gas-to-dust ratio obtained with different optical constants can vary up to a factor of $\approx1.4-1.9$, for the dustiest stars, and up to $\approx7$, in the most extreme cases, for dust poor sources.

The large range of possible values of $\Psi_{\rm dust}$ suggests that the choice of a unique value for $\Psi_{\rm dust}$, as usually assumed, might not in general be suitable for all the C-rich stars, as also noticed by \citet{Eriksson14}. 

The gas-to-dust ratio anti-correlates with the mass-loss rates. Such a trend suggest that the dust condensation efficiency increases with the density in the CSEs.
The relation between $\Psi_{\rm dust}$ and $\dot{M}$ is very steep for $\log\dot{M}\lessapprox-6$ and it tends to saturate for $\log\dot{M}\gtrapprox-6$. Indeed, for mass-loss rates larger than this threshold value, the dependence between $\Psi_{\rm dust}$ and $\dot{M}$ is milder.

\item \textit{Grain sizes}. For all the carbon dust optical data sets considered, larger grain sizes are predicted for redder stars. The most important parameter affecting the final grain size is the seed particle abundance.
The scaling relation between these two quantities is $a_{\rm amC}\propto(\epsilon_{\rm s})^{-1/3}$, as roughly recovered in our analysis.

\item \textit{Outflow expansion velocity}. The final velocity of the outflow is dependent on the optical data set assumed for carbon dust and on the seed particle abundance. Expansion velocities are expected to increase as a function of the mass-loss rate and of the luminosity. 
Above a certain value of $\dot{M}$, which depends on the set of optical constants, the outflow is accelerated. For $\dot{M}$ above this threshold, the velocity increases with the mass-loss rate, attaining its maximum value around $-5.7\lessapprox\dot{M}\lessapprox-5.6$. Typical values of the maximum velocity attained are between $20\lessapprox v_{\rm exp}\lessapprox25$~km~s$^{-1}$
However, the predicted velocities for the dustiest stars show a large scatter.

The velocities of the stars analyzed linearly scales with the stellar luminosity and with the gas-to-dust ratio, as predicted by the wind theory. For all the optical data sets, the outflow is accelerated above $L\approx2000$~$L_{\odot}$.

The velocities of the dustiest stars are typically $\approx1.9-2.7$ times larger than the ones adopted by \citet{Boyer12} and \citet{Srinivasan16} and $\approx1.1-1.7$ times larger than the value of $v_{\rm exp}=10$~km~s$^{-1}$, assumed by \citet{Groenewegen07}.

The trends between $v_{\rm exp}$ and $\dot{M}$ and between $v_{\rm exp}$ and $L$ are in general comparable to the ones observed for Galactic stars.
This result suggests that the dynamical properties of the outflow for carbon stars is rather independent of the metallicity. 
Such a trend is in agreement with the predictions of hydrodynamical calculations according to which the critical parameter determining the outflow expansion velocity is the carbon-excess, rather than the metallicity \citep{Mattsson10,Eriksson14}.

\item \textit{Other stellar quantities}. Input stellar quantities such as the carbon excess and the current stellar mass are not constrained by our analysis. Some degeneracy is expected among some of our models.
\end{itemize}

From the SED fitting analysis we compute the DPRs of the selected stars for the different optical data sets selected. We compare the results with the ones of the literature.
The main results are summarized below.
\begin{itemize}
\item The bulk of the dust produced is made of solid carbon for all the classes of stars. 
The amount of SiC dust can be up to $\approx8\%$ of the total. 
The amount of metallic iron produced is always negligible.
\item The total DPR is dominated by X-stars for which the contribution to the dust budget in mass is $\approx85\%$.
The DPRs obtained for different optical data sets of carbon dust are comparable within the estimated uncertainties, except for J1000 data set for which the DPR is lower. The variation of the total DPR computed with different optical constants is up to a factor $\approx2.4$.
\item Our DPR is in good agreement with the one derived by \citet{Matsuura13} for comparable optical constants of carbon dust.
However, the DPR estimated by these authors can be up to a factor $\approx2.4$ times larger than the 
the one derived by our analysis (see J1000). On the other hand, none among our data sets yields a DPR larger than the one by \citet{Matsuura13}.

The DPRs of \citet{Boyer12} and \citet{Srinivasan16} are $\approx 3$ times lower than the one we obtain for comparable optical constants of carbon dust.
Moreover, the largest and the lowest DPRs predicted by our analysis are respectively $\approx5$ and $\approx 2$ times larger than the ones by these authors. The difference found with these works for similar optical data sets can be due to several factors. For example, the predicted expansion velocities of our models are larger than the ones they adopted. Part of the discrepancy could also be ascribed to the assumptions related to the grain size distributions and to differences in the condensation radii and dust temperatures.
\end{itemize}

Our grids of models, including the spectra and other relevant dust and stellar quantities are publicly available at \url{http://starkey.astro.unipd.it/web/guest/dustymodels}.  

\subsection*{Acknowledgements}
This work is supported by the ERC Consolidator Grant funding scheme
({\em project STARKEY}, G.A. n.~615604).

\bibliographystyle{mn2e/mn2e_new}
\bibliography{nanni}

\appendix
\section{Photometry selection}
We here discuss how the observed photometry is selected to perform the SED fitting.

We visually inspected the SED of C-rich stars in the catalog, finding that in some cases the photometric points of the optical bands ($UBVI$), and, more rarely, of some bands in the NIR, cannot be associated to the SED of an AGB star.
Some examples in which the optical photometry is at odd with the NIR bands one is shown in Fig.~\ref{opt_cfr}. The disagreement found between the optical and NIR photometry can be due to a mismatch of the sources in different catalog or to the presence of a binary system. 
The observed photometry in the visual bands which are not in agreement with the NIR bands has been removed from our fitting procedure. 
For one particular source (SSTISAGEMAJ011219.72-735125.9), plotted in the upper panel of Fig.~\ref{opt_cfr}, we also exclude the $J$ and $H$ bands from the SED fitting procedure.
The list of sources for which some of the optical bands have been excluded from the SED fitting procedure are listed in Table~\ref{excluded}.

The photometry for the AKARI S11, L15 and WISE W3 is available for a sample of 117 sources in the catalog by \citet{Srinivasan16}.
For W3 a systematic offset towards fluxes lower than L15, which is around the same wavelength, is found by \citet{Srinivasan16}.
In Fig.~\ref{akari_opt_cfr} we show the fluxes obtained from the photometry listed in \citet{Srinivasan16} (red diamonds) overplotted with the IRS spectra (black crosses) and with the corresponding fluxes convolved with the S11 and L15 filters (black traingles).
For comparison, we also plot the flux of the AKARI bands obtained from the photometry in the catalog by \citet{Ita10}, indicated with blue asterisks. For the stars shown, \citet{Ita10} only contains data in the S11 filter.
From the two panels we notice that there is inconsistency between the photometry listed in \citet{Srinivasan16} and \citet{Ita10}. For the reasons discussed above, we exclude from the SED fitting procedure the AKARI S11, L15 and the WISE W3 fluxes. The exclusion of these photometric points is expected to produce only a minor effect on the results of the SED fitting.

For the MIR bands we consider in the fit the IRAC photometry at 4.5 and 5.8~$\mu$m only for dust enshrouded stars with $\jks\ge 2$.
In fact, as discussed in \citet{Nanni16}, the 4.5 and 5.8 bands of C-stars with a low amount of dust are affected by the C$_3$ feature, for which the available molecular opacity does not fit the feature. 

Another star (J012606.02-720921.0) is instead characterized by an observed photometry that cannot be reproduced by a SED from an AGB star, as already concluded by the fitting procedure of \citet{Srinivasan16}.

We exclude from the fit the photometry at 24~$\mu$m for all the FIR sources, since by definition FIR sources are more luminous at 24~$\mu$m than at 8~$\mu$m and the 24~$\mu$m flux cannot be fitted by the SED of an AGB.
The additional optical and NIR photometry excluded from the fitting procedure is listed in Table~\ref{excluded}.

For one additional source photometrically classified as C-rich (J011032.12-714146.3) the photometry at 24~$\mu$m appears to be unrelated to the SED.

\begin{figure}
\includegraphics[trim=0 0 0 -1cm, angle=90, width=0.48\textwidth]{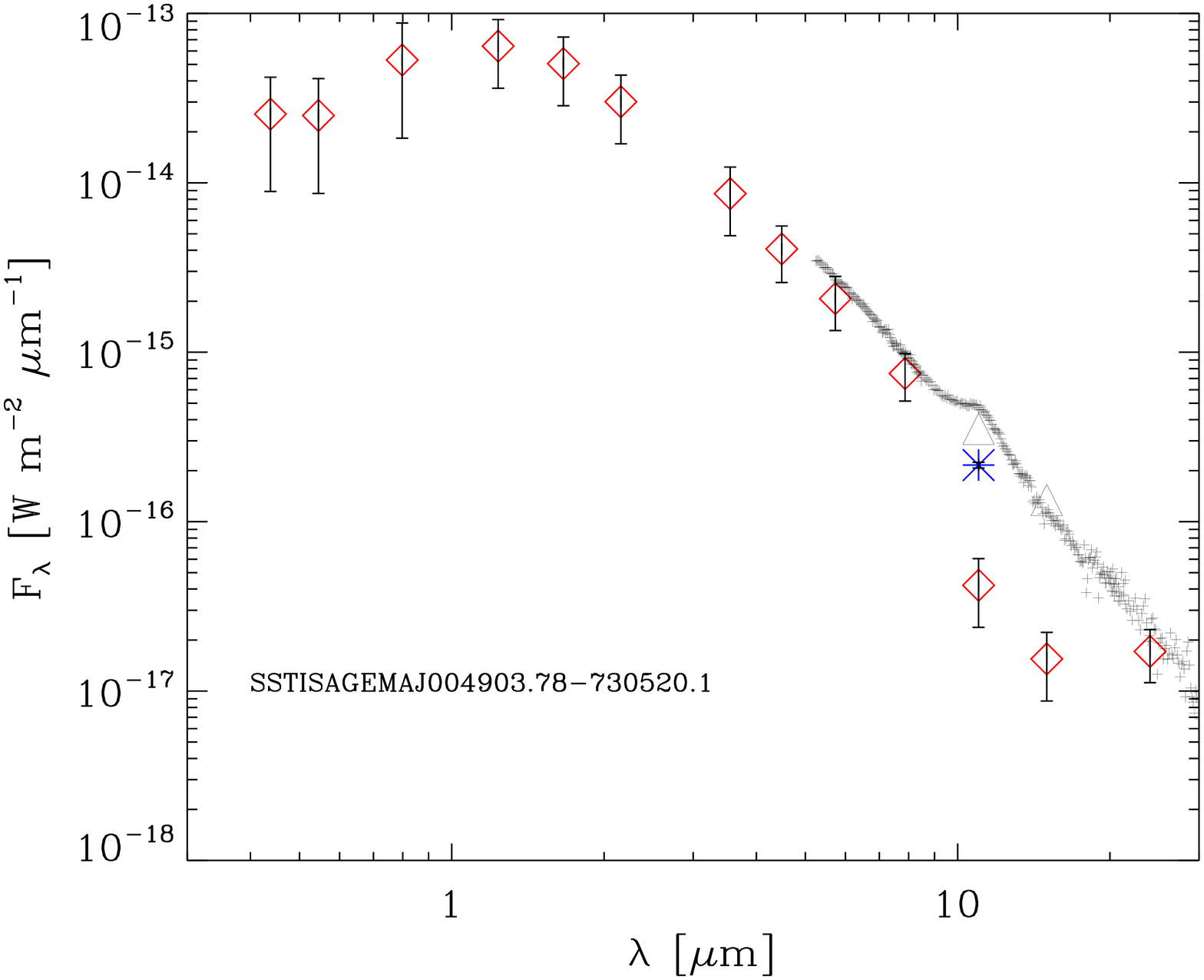}
\includegraphics[trim=0 0 0 -1cm, angle=90, width=0.48\textwidth]{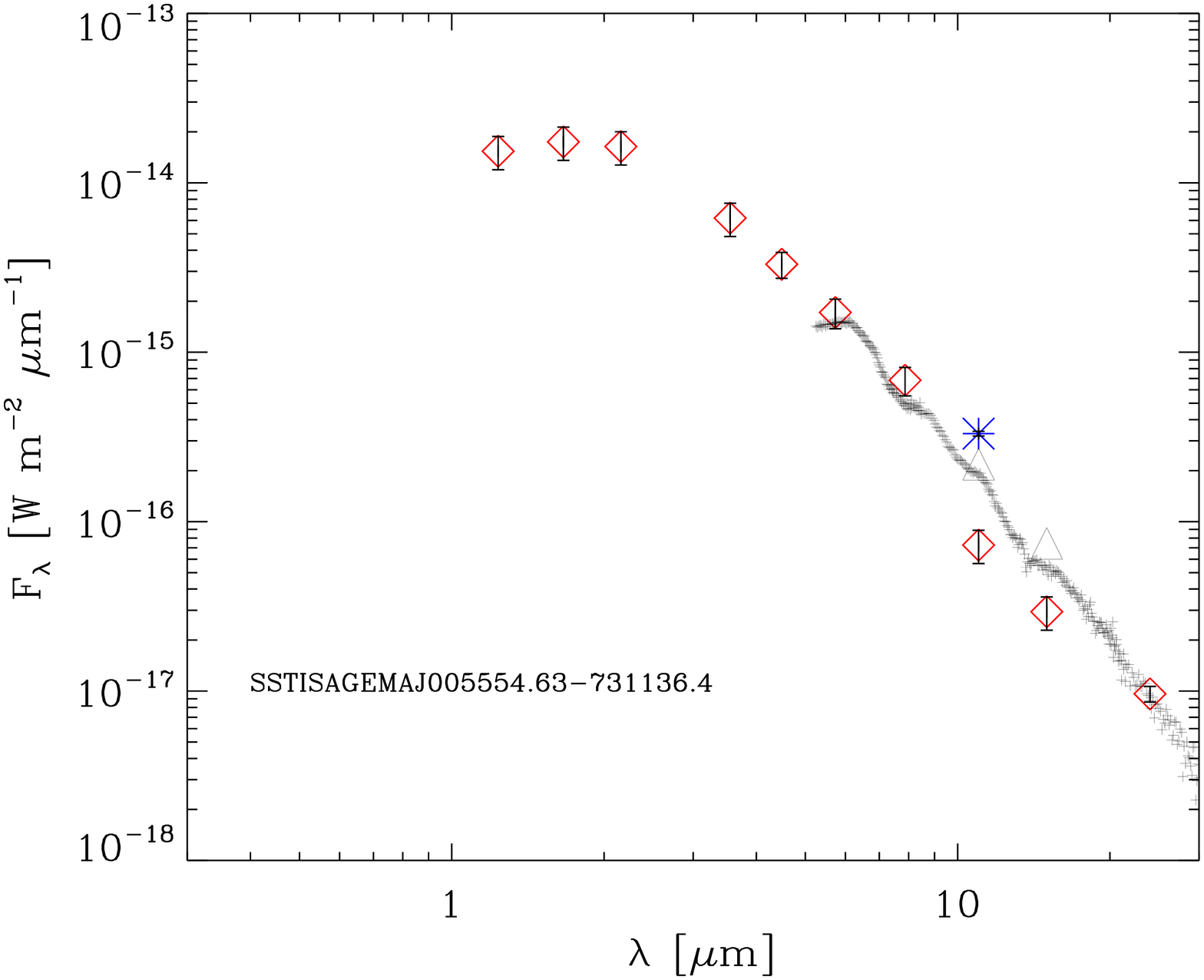}
        \caption{Two examples of sources showing the photometric data points taken from \citet{Srinivasan16} (red diamonds) and from \citet{Ita10} (blue asterisks). The AKARI S11 and L15 fluxes, computed convolving the filters with the observed IRS spectra, are represented by black triangles. The IRS spectra are also plotted with black crosses.}
        \label{akari_opt_cfr}
        \end{figure}

\begin{figure}
\includegraphics[trim=0 0 0 -1cm, angle=90, width=0.48\textwidth]{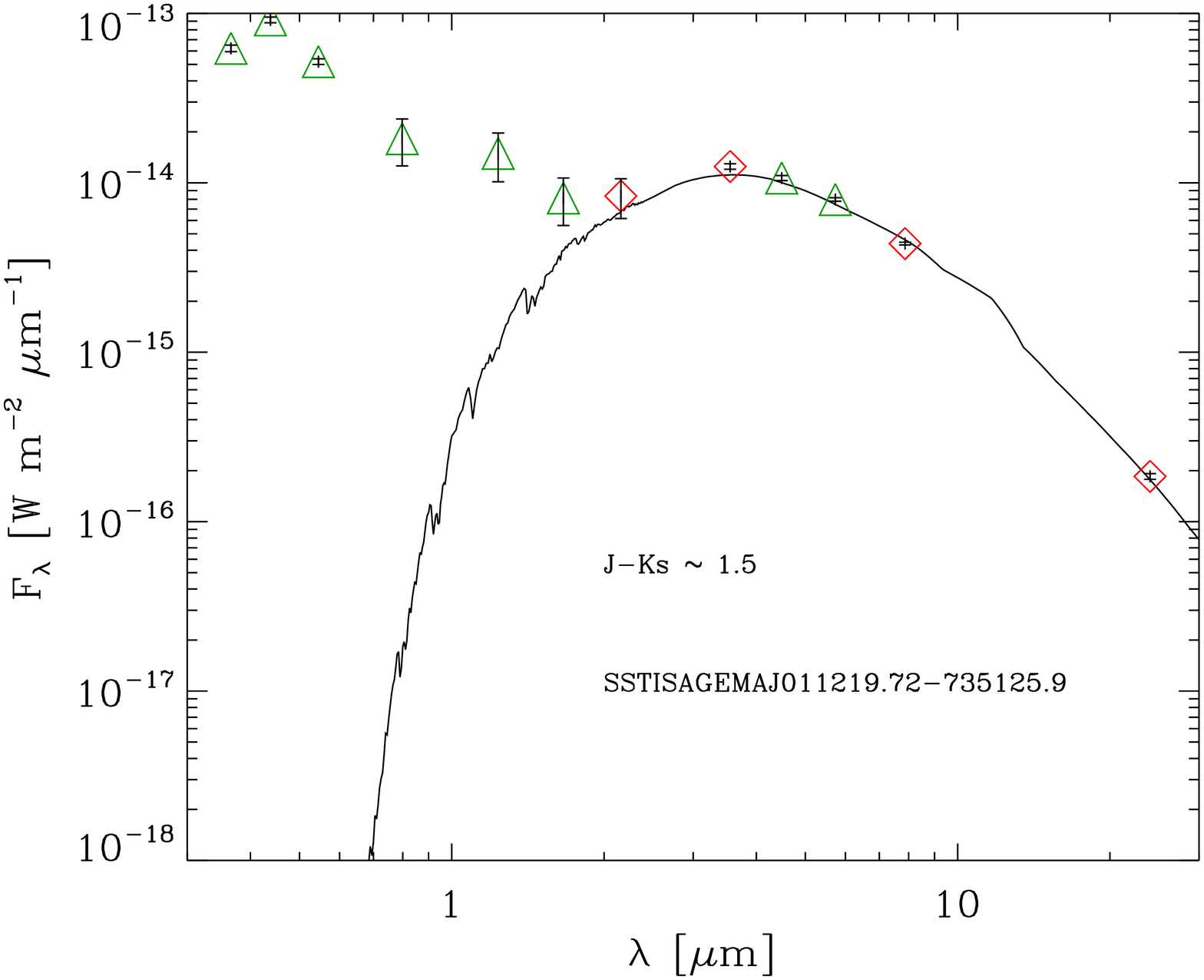}
\includegraphics[trim=0 0 0 -1cm, angle=90, width=0.48\textwidth]{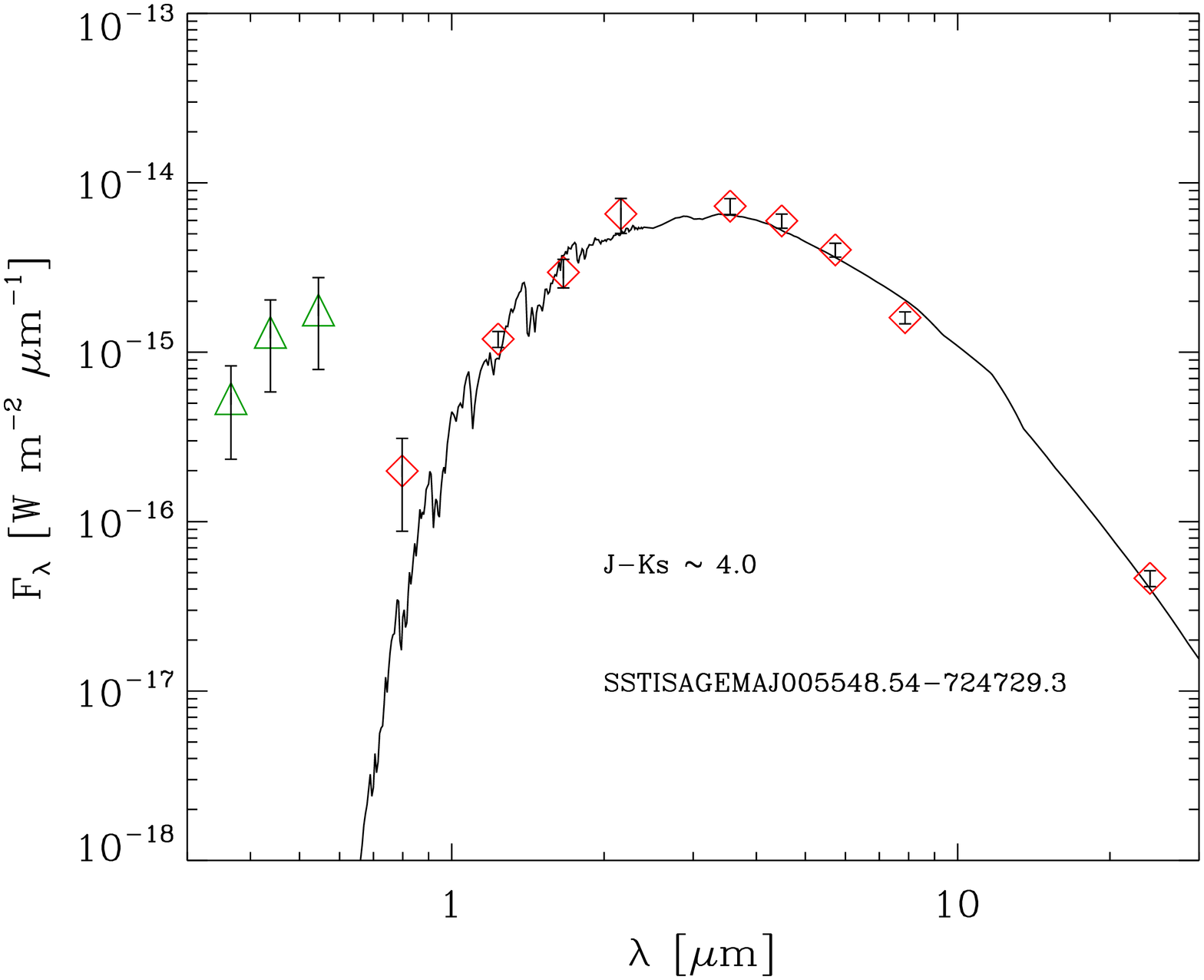}
\caption{Two examples of sources for which the optical photometric data points are not in agreement with the NIR ones. The observed photometric points, taken from the catalog by \citet{Srinivasan16}, are plotted with red diamonds, if the points are included in the fit, and with green triangles, if they are excluded. The uncertainty bars are also over plotted. The best fitting spectra among our grid of models are over plotted with a black solid line for each of the sources.}
        \label{opt_cfr}
        \end{figure}
        
\begin{table*}
\caption{Optical/NIR phometry excluded for the SED fitting.}
\label{excluded}
\begin{tabular}{c c}
\hline
Name & Optical/NIR bands excluded \\
\hline
SSTISAGEMAJ002124.62-734450.1& V I\\
SSTISAGEMAJ002358.52-733754.9& I\\
SSTISAGEMAJ002942.24-731911.1& U\\
SSTISAGEMAJ003040.75-734245.1& B V\\
SSTISAGEMAJ003117.69-722726.0& U\\
SSTISAGEMAJ003314.99-740211.7& U\\
SSTISAGEMAJ003346.13-724458.5& B V\\
SSTISAGEMAJ003639.49-721626.6& B V\\
SSTISAGEMAJ003905.71-724716.3& B\\
SSTISAGEMAJ004116.96-725216.6& U\\
SSTISAGEMAJ004125.96-735345.6& U\\
SSTISAGEMAJ004210.64-735003.6& B\\
SSTISAGEMAJ004221.80-722405.8& U B V\\
SSTISAGEMAJ004305.88-732140.7& V\\
SSTISAGEMAJ004308.75-734825.2& B\\
SSTISAGEMAJ004325.20-721851.0& B V\\
SSTISAGEMAJ004333.05-724803.7& U\\
SSTISAGEMAJ004339.55-731457.6& U B\\
SSTISAGEMAJ004346.41-733828.0& B\\
SSTISAGEMAJ004353.49-724854.5& B V I\\
SSTISAGEMAJ004442.81-725914.6& U B\\
SSTISAGEMAJ004453.12-721606.0& B\\
SSTISAGEMAJ004457.03-730556.0& U\\
SSTISAGEMAJ004502.13-725224.2& B V\\
SSTISAGEMAJ004516.87-734133.4& U\\
SSTISAGEMAJ004641.37-730613.5 & U B V I J\\
SSTISAGEMAJ004648.02-731709.4& B\\
SSTISAGEMAJ004657.56-723224.8& U\\
SSTISAGEMAJ004720.41-722503.7& U\\
SSTISAGEMAJ004756.89-722614.5& U\\
SSTISAGEMAJ004805.77-731743.5 & B V I\\
SSTISAGEMAJ004825.72-724402.8& V I\\
SSTISAGEMAJ004903.78-730520.1& B\\
SSTISAGEMAJ004916.97-732229.0& U B\\
SSTISAGEMAJ004922.69-731454.5& U\\
SSTISAGEMAJ004931.48-730715.9& B\\
SSTISAGEMAJ004941.35-731623.8& U B\\
SSTISAGEMAJ004942.70-730220.8& B V\\
SSTISAGEMAJ004949.95-724514.4& V I\\
SSTISAGEMAJ004956.41-725933.4& U B\\
SSTISAGEMAJ005012.17-725319.0& U B\\
SSTISAGEMAJ005013.29-731113.1& U B\\
SSTISAGEMAJ005015.26-731243.8& U B\\
SSTISAGEMAJ005016.63-732517.7& B\\
SSTISAGEMAJ005022.85-732249.7& U B\\
SSTISAGEMAJ005031.30-722913.0& B\\
SSTISAGEMAJ005045.34-731839.6& U\\
SSTISAGEMAJ005052.64-725216.5& B V\\
SSTISAGEMAJ005057.36-730106.1& U B\\
SSTISAGEMAJ005102.05-725925.6& B\\
SSTISAGEMAJ005105.61-733625.0& B V I\\
SSTISAGEMAJ005109.24-731933.3& B\\
SSTISAGEMAJ005145.14-732033.7& U B\\
SSTISAGEMAJ005145.18-731805.3& U\\
SSTISAGEMAJ005148.79-730243.8& U B\\
\hline
\end{tabular}
\begin{tabular}{c c}
\hline
Name & Optical/NIR bands excluded\\
\hline
SSTISAGEMAJ005207.42-732138.2& U B\\
SSTISAGEMAJ005222.67-721846.0& U\\
SSTISAGEMAJ005224.60-721115.2& U B\\
SSTISAGEMAJ005229.78-734657.3& U\\
SSTISAGEMAJ005233.39-725409.8& U B\\
SSTISAGEMAJ005259.03-722811.2& U B\\
SSTISAGEMAJ005313.98-731517.4& U\\
SSTISAGEMAJ005315.62-731212.9& U\\
SSTISAGEMAJ005354.59-732204.0& B V I\\
SSTISAGEMAJ005356.22-703804.1& U B V I\\
SSTISAGEMAJ005357.32-733433.0& B\\
SSTISAGEMAJ005407.65-723628.7& U B\\
SSTISAGEMAJ005408.52-721420.4& B\\
SSTISAGEMAJ005421.12-731544.3& B\\
SSTISAGEMAJ005422.26-724329.8& V I\\
SSTISAGEMAJ005433.07-725814.2& B\\
SSTISAGEMAJ005509.96-734244.7& B\\
SSTISAGEMAJ005514.27-732505.3& U B V I\\
SSTISAGEMAJ005526.63-724514.3& U\\
SSTISAGEMAJ005528.97-724848.6& U B\\
SSTISAGEMAJ005533.34-724127.1& B\\
SSTISAGEMAJ005548.54-724729.3& U B V\\
SSTISAGEMAJ005625.52-740047.0& B V\\
SSTISAGEMAJ005656.43-733530.1& B V\\
SSTISAGEMAJ005705.82-741316.5& U\\
SSTISAGEMAJ005710.93-723100.1& U B V\\
SSTISAGEMAJ005727.64-725327.9& U B V\\
SSTISAGEMAJ005749.03-730521.5& U B\\
SSTISAGEMAJ005858.17-723911.5& U\\
SSTISAGEMAJ005936.59-722717.0& U\\
SSTISAGEMAJ005942.96-730443.2& U B\\
SSTISAGEMAJ005958.80-720300.9& B\\
SSTISAGEMAJ010015.66-722226.1& U\\
SSTISAGEMAJ010053.23-722333.6& U\\
SSTISAGEMAJ010149.32-724900.0& U\\
SSTISAGEMAJ010218.12-722820.8& U\\
SSTISAGEMAJ010245.16-741257.4& U B V I\\
SSTISAGEMAJ010307.18-720629.5& B V I J\\
SSTISAGEMAJ010441.50-712609.9& I\\
SSTISAGEMAJ010442.35-730142.8& U\\
SSTISAGEMAJ010503.13-715929.7& U B V I J H\\
SSTISAGEMAJ010552.88-721902.4& U\\
SSTISAGEMAJ010555.92-723244.0& U B\\
SSTISAGEMAJ010615.16-721653.1& U\\
SSTISAGEMAJ010752.86-714617.1& B\\
SSTISAGEMAJ010810.35-725307.9& B\\
SSTISAGEMAJ010812.96-725243.9& U B V\\
SSTISAGEMAJ010820.66-725252.0& U B\\
SSTISAGEMAJ010858.23-724142.1& B\\
SSTISAGEMAJ010901.19-732015.3& U\\
SSTISAGEMAJ010928.94-722821.2& B\\
SSTISAGEMAJ011032.24-730504.0& B V\\
SSTISAGEMAJ011219.65-724108.4& V I\\
SSTISAGEMAJ011219.72-735125.9& U B V I J H\\
SSTISAGEMAJ011642.41-722725.7& U\\
\hline
\end{tabular}
\end{table*}

\label{lastpage}
\end{document}